%% file: PeriodicHOSkipping.tex
\documentclass[12pt,draftclsnofoot,onecolumn]%
              {IEEEtran}
\usepackage[T1]{fontenc}
\ifCLASSINFOpdf
  \usepackage[pdftex]{graphicx,color,hyperref}
\else
  \usepackage[dvipdfmx,dvips]{graphicx,color,hyperref}
\fi
\usepackage{tikz}
\usepackage[cmex10]{amsmath}
\usepackage[cmintegrals]{newtxmath}
\usepackage{bm,cite,enumerate,url}
\usepackage[caption=false,font=footnotesize]{subfig}
\newcommand{\Prb}{\mathbb{P}}\newcommand{\Exp}{\mathbb{E}}

\newcommand{\dd}{\mathrm{d}}
\newcommand{\R}{\mathbb{R}}
\newcommand{\N}{\mathbb{N}}
\newcommand{\B}{\mathcal{B}}

\newcommand{\SINR}{\mathsf{SINR}}

\let\disp\displaystyle
\let\Bar\overline\let\Tilde\widetilde

\newtheorem{proposition}{Proposition}
\newtheorem{theorem}{Theorem}
\newtheorem{corollary}{Corollary}
\newtheorem{lemma}{Lemma}
\newtheorem{remark}{Remark}

\begin{document}\sloppy\allowbreak\allowdisplaybreaks

\title{%
  Periodic Handover Skipping in\\ Cellular Networks: Spatially
  Stochastic Modeling and Analysis%
  \thanks{This work has been presented in part in~\cite{TokuMiyo18}.}
}  
\author{%
  Kiichi Tokuyama, Tatsuaki Kimura, and Naoto Miyoshi%
  \thanks{K. Tokuyama and N. Miyoshi are with the Department of
    Mathematical and Computing Science, Tokyo Institute of
    Technology, Tokyo, Japan.
    E-mail:
    \href{mailto:tokuyama.k.aa@m.titech.ac.jp}{tokuyama.k.aa@m.titech.ac.jp},
    and
    \href{mailto:miyoshi@is.titech.ac.jp}{miyoshi@is.titech.ac.jp}.
    T. Kimura is with the Department of Information and Communications
    Technology, Osaka University, Osaka, Japan.
    E-mail:
    \href{mailto:kimura@comm.eng.osaka-u.ac.jp}{kimura@comm.eng.osaka-u.ac.jp}.
  }%
  \thanks{This work was supported by the Japan Society for the
    Promotion of Science (JSPS) Grant-in-Aid for Scientific Research
    (C) 19K11838.
  }%
}
    
\maketitle
%\IEEEpeerreviewmaketitle

\begin{abstract}
Handover (HO) management is one of the most crucial tasks in dense
cellular networks with mobile users.
A problem in the HO management is to deal with increasing HOs due to
network densification in the 5G evolution and various HO skipping
techniques have so far been studied in the literature to suppress
excessive HOs.
In this paper, we propose yet another HO skipping scheme, called
\textit{periodic HO skipping}.
The proposed scheme prohibits the HOs of a mobile user equipment~(UE)
for a certain period of time, referred to as skipping period, thereby
enabling flexible operation of the HO skipping by adjusting the length
of the skipping period.
We investigate the performance of the proposed scheme on the basis of
stochastic geometry.
Specifically, we derive analytical expressions of two performance
metrics---the HO rate and the expected downlink data rate---when a UE
adopts the periodic HO skipping.
Numerical results based on the analysis demonstrate that the periodic
HO skipping scenario can outperform the scenario without any HO
skipping in terms of a certain utility metric representing the
trade-off between the HO rate and the expected downlink data rate, in
particular when the UE moves fast.
Furthermore, we numerically show that there can exist an optimal
length of the skipping period, which locally maximizes the utility
metric, and approximately provide the optimal skipping period in a
simple form.
Numerical comparison with some other HO skipping techniques is also
conducted.
%
%Handover (HO) management is one of the most crucial tasks in dense
%cellular networks with mobile users.  A problem in the HO management
%is to deal with increasing HOs due to network densification in the 5G
%evolution and various HO skipping techniques have so far been studied
%in the literature to suppress excessive HOs.  In this paper, we
%propose yet another HO skipping scheme, called periodic HO skipping.
%The proposed scheme prohibits the HOs of a mobile user equipment (UE)
%for a certain period of time, referred to as skipping period.  We
%investigate the performance of the proposed scheme on the basis of
%stochastic geometry.  Specifically, we derive analytical expressions
%of two performance metrics---the HO rate and the expected downlink
%data rate---when a UE adopts the periodic HO skipping.  Numerical
%results based on the analysis demonstrate that the periodic HO
%skipping scenario can outperform the scenario without any HO skipping
%in terms of a certain utility metric representing the trade-off
%between the HO rate and the expected downlink data rate, in particular
%when the UE moves fast.  Furthermore, we numerically show that there
%can exist an optimal length of the skipping period which locally
%maximizes the utility metric and approximately provide the optimal
%skipping period in a simple form.
\end{abstract}

\section{Introduction}\label{sec:Intro}

The development of the fifth generation mobile communication
systems~(5G) is driven by the ever-increasing demand for channel
capacity due to the proliferation of mobile user equipments~(UEs) such
as mobile phones, tablets, and other handheld devices.
One of the key solutions in the 5G evolution is network densification
through small cell deployments~(see, e.g., \cite{6736747,7476821}).
Densifying base stations (BSs) offers more capacity, which improves
the quality of service.
On the other hand, it shrinks the service area of each BS and induces
frequent handovers (HOs), which may increase the signaling overhead
and the risk of disconnections.

HO skipping is an approach to address the problem of frequent HOs by
skipping some opportunities of HOs (see, e.g.,
\cite{ArshElSaSoroAlNaAlou16,ArshElSaSoroAlNaAlou17,ArshElSaSoroAlNaAlou16b,ArshElSaSoroAlNaAlou16c,DemaPsomKrik18,8662599,8885814}).
However, in turn, the HO skipping may decrease the data reception rate
(data rate for short) since it tends to force a UE to retain
long-distance connection with a BS.
In other words, the HO skipping induces a trade-off between the HO
rate and the data rate, and this trade-off should be balanced for the
network densification to work effectively.
While various HO skipping techniques have so far been proposed and
studied in such a point of view, we propose yet another HO skipping
scheme, called \textit{periodic HO skipping}.
The proposed scheme prohibits the HOs of a mobile UE for a certain
period of time, referred to as \textit{skipping period}, thereby
enabling flexible operation of the HO skipping by adjusting the length
of the skipping period.
In this paper, we investigate the performance of the proposed scheme
from the perspective of the trade-off between the HO rate and the data
rate.

\subsection{Related Work}

A number of studies have so far analyzed the performance of cellular
networks with mobile UEs and many of them have adopted stochastic
geometry as an analytical tool (see, e.g., recent tutorial
articles~\cite{ElSaSultAlouWin17,TabaSaleHoss19} and references
therein).
In the stochastic geometry approach, the locations of wireless nodes
(BSs and/or UEs) in a wireless network are modeled by stochastic point
processes on the Euclidean plane, so that we can capture the spatial
irregularity of wireless nodes and explore mathematical analysis of
region-independent network performance by virtue of the theory of
point processes and stochastic geometry.
The first results along this line date back to the late
1990s~\cite{BaccZuye97,BaccKleiLeboZuye97}, where the cells associated
with BSs in a cellular network are modeled as the Voronoi tessellation
formed by a homogeneous Poisson point process~(PPP) and some
performance metrics concerning mobile UEs are discussed.
Since the 2010s, this stream has become more active.
Lin~\textit{et al.}~\cite{LinGantFlemAndr13} propose a mobility model
of a UE on single-tier hexagonal/PPP networks and analyze the HO rate
and the expected sojourn time of a mobile UE staying in a particular
cell.
The results in \cite{LinGantFlemAndr13} are then extended in
\cite{HongXuTaoLiSven15} to a two-tier heterogeneous network~(HetNet).
Bao and Liang~\cite{BaoLian15} derive an analytical expression for the
HO rate in a multi-tier HetNet modeled using overlaid independent PPPs
and provide a guideline for tier selection taking both the HO rate and
the expected downlink data rate into account.
In addition, \cite{BaoLian16} develops a similar analysis to
\cite{BaoLian15} for a single-tier network with BS cooperation.
Sadr and Adve~\cite{SadrAdve15} analyze the HO rate in a PPP model of
multi-tier HetNets with orthogonal spectrum allocation among tiers and
investigate the negative impact of HOs on the coverage probability.
Chattopadhyay~\textit{et al.}~\cite{ChatBlasAltm19} evaluate the
expected downlink data rate for a mobile UE taking into account the
data outage periods due to HOs in a two-tier HetNet and further
discuss the fraction of connecting BSs to reduce frequent HOs.

Several HO skipping techniques have also been proposed and analyzed
using the stochastic geometry (see, e.g.,
\cite{ArshElSaSoroAlNaAlou16,ArshElSaSoroAlNaAlou17,ArshElSaSoroAlNaAlou16b,ArshElSaSoroAlNaAlou16c,DemaPsomKrik18}).
Arshad~\textit{et al.}~\cite{ArshElSaSoroAlNaAlou16} introduce the
so-called alternate HO skipping, where a mobile UE executes HOs
alternately along its trajectory, and quantify the average throughput
representing the trade-off between the HO rate and the expected
downlink data rate when a UE adopts the alternate HO skipping.
The results in \cite{ArshElSaSoroAlNaAlou16} are extended in
\cite{ArshElSaSoroAlNaAlou17} to a two-tier HetNet and in
\cite{ArshElSaSoroAlNaAlou16b} to the incorporation with BS
cooperation.
Furthermore, \cite{ArshElSaSoroAlNaAlou16c} proposes topology-aware HO
skipping, where a UE skips an HO when the target BS is far from the
UE's trajectory or the cell of the target BS is small, and evaluates
its performance by Monte Carlo simulations.
Demarchou~\textit{et al.}~\cite{DemaPsomKrik18} then provide a
mathematical analysis of the topology-aware HO skipping.
Compared with these sophisticated HO skipping techniques, our proposed
scheme is simple and easy to implement since it is enough for a mobile
UE to observe the BS locations every fixed-length period (as seen in
the definition of the scheme in Sec.~\ref{subsec:HOSkipping}).
We will find later that such a simple scheme can compete with the
sophisticated ones.

\subsection{Contributions}

The contributions of this work are summarized as follows.
\begin{enumerate}
\item\label{cont1} We propose and advocate the periodic HO skipping,
  which prohibits the HOs of a mobile UE during each cycle of the
  skipping period.
\item\label{cont2} Applying the stochastic geometry approach, we
  derive the analytical expressions of the HO rate and the expected
  downlink data rate when a mobile UE adopts the periodic HO skipping.
\item\label{cont3} On the basis of the analytical results, we
  numerically demonstrate that the proposed scheme can outperform the
  conventional scenario without any HO skipping, in particular when
  the UE moves fast.
\item\label{cont4} We numerically observe that there can exist an
  optimal length of the skipping period and provide an approximate
  optimal skipping period in a simple computable form.
\item\label{cont5} We numerically observe that the proposed scheme can
  compete with some other sophisticated HO skipping techniques.
\end{enumerate}
This work enhances \cite{TokuMiyo18}, where the periodic HO skipping
is already proposed.
However, the contributions in analysis~\ref{cont2}) and the numerical
experiments~\ref{cont3}) are fundamentally refined.
Moreover, the contributions~\ref{cont4}) and~\ref{cont5}) are
completely new.

\subsection{Organization}

The rest of the paper is organized as follows.
In the next section, we describe the network model and our proposed
periodic HO skipping scheme.
We then define the user mobility model and the performance metrics;
that is, the expected downlink data rate and the HO rate.
In Sec.~\ref{sec:analysis}, the performance of the proposed scheme is
investigated, where for comparison, we analyze the performance metrics
not only for the proposed scheme but also for the scenario without any
HO skipping.
Then, on the basis of the analytical results, the performances of the
two scenarios are numerically compared in terms of a certain utility
metric representing the trade-off between the data rate and the HO
rate.
In Sec.~\ref{sec:s_opt}, we discuss how to decide the length of the
skipping period, where we numerically observe that there exists an
optimal length of the skipping period which locally maximizes the
utility metric.
We then provide a simple computable expression of an approximate
optimal skipping period.
Some properties of the approximate optimal skipping period are also
revealed by numerical experiments.
Numerical comparison with some other HO skipping techniques are made
in Sec.~\ref{sec:comparison}.
Finally, the paper is concluded in Sec.~\ref{sec:conclusion}.

\section{System Model and Periodic Handover Skipping}\label{sec:model}

\subsection{Network Model}\label{subsec:Network}

In this paper, we develop our proposed periodic HO skipping scheme
implemented on the most basic spatially stochastic model of cellular
networks; that is, a homogeneous PPP network with Rayleigh fading and
power-law path-loss (see, e.g.,
\cite{AndrBaccGant11,BlasHaenKeelMukh18}).
Before defining the proposed scheme, we detail here the network
model.

BSs in a cellular network are deployed according to a homogeneous
PPP $\Phi = \sum_{i\in\N}\delta_{X_i}$ on the Euclidean plane $\R^2$
with intensity~$\lambda\in(0,\infty)$, where $\N := \{1,2,\ldots\}$,
$\delta_x$ denotes the Dirac measure with mass at $x\in\R^2$ and the
points~$X_1, X_2, \ldots$ of $\Phi$ are numbered in an arbitrary
order.
All the BSs transmit signals with the same power level, normalized to
one, using a common spectrum bandwidth.
We suppose that the time is divided into discrete slots and the
downlink channels are affected by Rayleigh fading and power-law
path-loss, whereas shadowing effects are ignored.
Therefore, if a UE located at $\bm{u}\in\R^2$ at slot~$t\in\N_0 :=
\N\cup\{0\}$ receives a signal from the BS at $X_i$, $i\in\N$, the
received signal power is represented by $H_{i,t}\|X_i -
\bm{u}\|^{-\beta}$, where $\|\cdot\|$ stands for the Euclidean norm,
$H_{i,t}$, $i\in\N$, $t\in\N_0$, are mutually independent and
exponentially distributed random variables with unit mean representing
the fading effects, and $\beta>2$ denotes the path-loss exponent.

We assume that, at any time slot, each BS has at least one UE in
service and transmits a signal to one of its UEs.
Then, if a UE is located at $\bm{u}\in\R^2$ at slot~$t\in\N_0$ and is
served by the BS at $X_i$, the downlink
signal-to-interference-plus-noise ratio (SINR) for this UE is
represented by
\begin{equation}\label{eq:SINR}
  \SINR_{\bm{u},i}(t)
  = \frac{H_{i,t}\|X_i - \bm{u}\|^{-\beta}}
         {I_{\bm{u},i}(t) + \sigma^2},
  \quad i\in\N,\; t\in\N_0,
\end{equation}
where $\sigma^2$ denotes a nonnegative constant representing the noise
power and $I_{\bm{u},i}(t)$ denotes the total interference power to
this UE given by
\begin{equation}\label{eq:Interference}
  I_{\bm{u},i}(t)
  = \sum_{j\in\N\setminus\{i\}}
      H_{j,t}\,\|X_j - \bm{u}\|^{-\beta}.
\end{equation}
The instantaneous downlink data rate~$\xi_{\bm{u},i}(t)$ is then
defined as
\begin{equation}\label{eq:DataRate}
  \xi_{\bm{u},i}(t)
  =  \log (1 + \SINR_{\bm{u},i}(t)),
\end{equation}  
where $\log$ stands for the natural logarithm for simplicity, but of
course, it can be converted into the conventional binary logarithm by
multiplying the constant $\log_2e$.

\subsection{Periodic Handover Skipping}\label{subsec:HOSkipping}

Suppose that a UE moves on $\R^2$ and is initially (at slot~$0$)
served by its nearest BS, which offers the strongest signal power to
the UE when the fading effects are averaged out.
In other words, the cells of respective BSs form a Poisson-Voronoi
tessellation~(see \cite[Sec.~9.7]{ChiuStoyKendMeck13}).
In our proposed periodic HO skipping scheme, a UE is prohibited from
executing HOs and retains the initial connection for $s$~time slots,
referred to as \textit{skipping period}, regardless of its motion.
After the skipping period of $s$~slots has passed, the UE reexamines
the connection and if the current connection is no longer with the
nearest BS due to its moving, the UE executes an HO and makes a new
connection with the nearest one.
Afterward, this procedure is repeated in cycles of the skipping
period.
Namely, the UE reexamines its connection to a BS every $s$~time slots,
during which it skips any chances of HOs even if it crosses the
boundaries between cells.
We assume that an HO, if it is done, is executed instantly without any
time loss.
We should note that a UE is always connected to its nearest BS at the
beginning of each cycle of the skipping period, whereas it does not
always execute an HO at the end of a cycle since the current
connection can still be with the nearest one.

One may claim that the skipping period described above is similar to
the time-to-trigger (TTT) in Long-Term Evolution
(LTE)~\cite{3gpp.36.331}.
Indeed, they both suppress the number of HOs by prohibiting them for a
certain period of time.
However, they are substantially different in that the TTT starts at
the instant that a UE crosses a boundary between two cells and it is
in the order of 100msec, which mainly aims to prevent ping-pong
phenomena around the boundaries between cells.
On the other hand, the skipping period repeats in cycles and prohibits
HOs during each cycle, which is in the order of seconds (though it may
depend on the speed of the UE).

Clearly, the choice of the length of the skipping period is vital for
our proposed scheme.
If the skipping period is too short, it results in frequent HOs,
whereas the long skipping period may cause long-distance connections,
which deteriorate the transmission performance.
Therefore, we should decide the length of the skipping period
carefully.

\subsection{Mobility Model}\label{subsec:Mobility}

\begin{figure}[!t]
  \centering  
  \includegraphics[width=.5\linewidth]{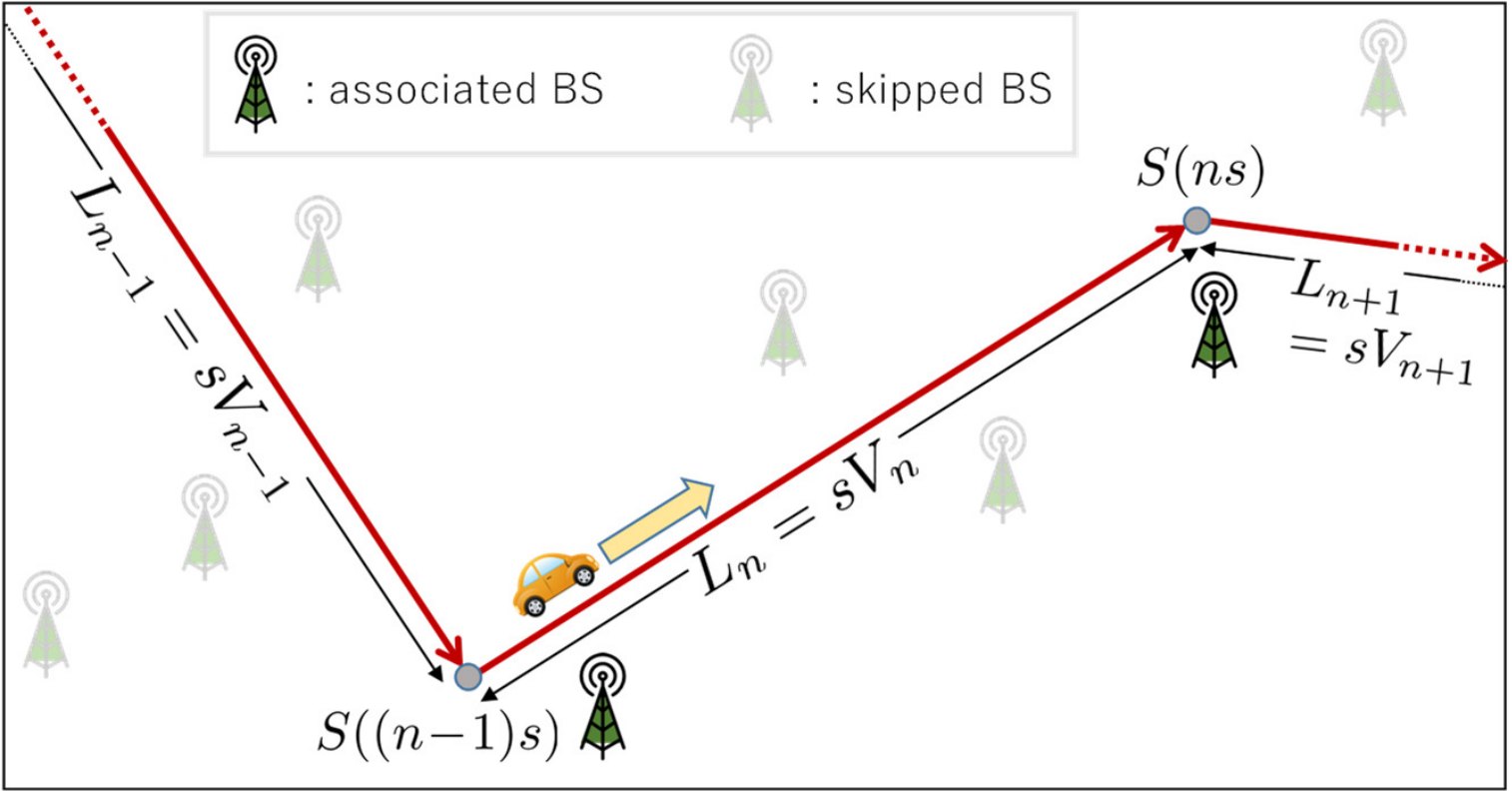}
  \caption{A path of the typical UE in the random walk mobility
    model.}\label{fig:path}
\end{figure}

Owing to the spatial stationarity of the network model, we can focus
on a UE that is supposed to be at the origin~$\bm{0}=(0,0)\in\R^2$ at
slot~$0$ and we refer to this UE as the \textit{typical UE}.
Let $S(t)$ denote the location of the typical UE at slot~$t\in\N_0$.
Since a UE is allowed to execute an HO every cycle of the skipping
period in our proposed scheme, it is enough to observe the location of
the typical UE every cycle and we model its motion as a simple and
tractable random walk on $\R^2$.
Let $Y_1,Y_2,\ldots$ denote a sequence of independent and identically
distributed (i.i.d.) random variables on $\R^2$ representing the
motions of the typical UE in respective cycles of the skipping period.
Then, the location of the typical UE just after $n$~cycles (that is,
at slot~$n s$) is provided as a random work;
\begin{equation}\label{eq:P_ns}
  S(n s) = \sum_{k=1}^n Y_k, \quad n\in\N,
\end{equation}
with $S(0) = \bm{0}$.
We assume that the typical UE moves along the straight line segment at
a constant velocity during each cycle; that is, $\{S(t)\}_{t\in\N_0}$
is piecewise deterministic and is given by
\begin{equation}\label{eq:P_t}
  S(t) = S(n s) + \frac{t - n s}{s}\,Y_{n+1},
  \quad t = n s, n s + 1, \ldots, (n+1)s,\; n\in\N_0.
\end{equation}
An example of a path of the typical UE is illustrated in
Fig.~\ref{fig:path}.
Let $Y_n = (L_n, \psi_n)$, $n\in\N$, in the polar coordinates.
Then, the moving speed of the typical UE during the $n$th cycle is
equal to $V_n = L_n/s$.
It is reasonable to suppose that the moving distance~$L_n$ in a cycle
depends on the cycle length~$s$; that is, $L_n$ is stochastically
larger as $s$ is larger.
Hence, we provide the distribution of $V_n$, instead of $L_n$, and
that of $\psi_n$ for our mobility model and assume that these
distributions do not depend on the cycle length.
The distributions of $V_n$ and $\psi_n$ respectively represent changes
in speed and direction of the typical UE over cycles of the skipping
period, and the choice of these distributions gives enough flexibility
to our model to capture various mobility patterns.
For instance, $\Prb(V_n = 0) > 0$ represents that the UE can take a
pause for $s$~time slots with a positive probability, and if $\psi_n$
takes a constant, the UE always moves along a straight line.

\subsection{Performance Metrics}

As discussed in Sec.~\ref{sec:Intro}, the HO skipping induces the
trade-off between the HO rate and the data rate.
We thus evaluate the performance of our proposed scheme in terms of
the expected downlink data rate~$\mathcal{T}$ and the HO
rate~$\mathcal{H}$, which are respectively defined as
\begin{align}
  \mathcal{T}
  &= \lim_{m\to\infty}
       \frac{1}{m}\Exp\biggl[\sum_{t=0}^{m-1}\xi(t)\biggr],
  \label{eq:T}\\
  \mathcal{H}
  &= \lim_{m\to\infty}\frac{\Exp[\zeta(0,m)]}{m},
  \label{eq:H}
\end{align}
where $\xi(t)$ denotes the instantaneous downlink data rate for the
typical UE at slot~$t$, specifically given in~\eqref{eq:DataRate}, and
$\zeta(a,b)$ denotes the number of HOs executed by the typical UE from
slot~$a$ to slot~$b$.
These performance metrics are analyzed and evaluated in the following
sections.

\section{Performance Analysis and Evaluation}\label{sec:analysis}

In this section, we investigate the performance of our proposed scheme
introduced in the preceding section.
For comparison, we analyze the performance metrics defined in
\eqref{eq:T} and \eqref{eq:H} not only in the scenario with the
proposed periodic HO skipping but also in the conventional scenario
without any HO skipping on the same network and mobility models
described in Secs.~\ref{subsec:Network} and \ref{subsec:Mobility}.
In the scenario without HO skipping, the typical UE certainly executes
an HO whenever it crosses a boundary between two cells.
We refer to the scenario without HO skipping and that with the
periodic HO skipping as Scenario~0 and Scenario~1, respectively, and
distinguish elements in the respective scenarios by putting the
superscript ``(0)'' or ``(1)''; for example, $\mathcal{T}^{(0)}$ and
$\mathcal{H}^{(0)}$ respectively stand for the expected downlink data
rate and the HO rate in Scenario~0, whereas $\mathcal{T}^{(1)}$ and
$\mathcal{H}^{(1)}$ are those in Scenario~1.

\subsection{Expected Downlink Data Rate Analysis}

\subsubsection{Expected Downlink Data Rate in Scenario~0}
In Scenario~0, the typical UE certainly executes an HO whenever it
crosses a boundary between two cells; that is, the typical UE is
always connected to its nearest BS.
The following proposition is directly derived from the existing result
in the literature.

\begin{proposition}\label{prp:DataRate0}
For the cellular network model described in Sec.~\ref{sec:model},
the expected downlink data rate in Scenario~0 is given by
\begin{equation}\label{eq:DataRate0}
  \mathcal{T}^{(0)}
  = \int_0^\infty\!\!\!\int_0^\infty
      \frac{\rho(z,w)}{1+z}\,
    \dd z\,\dd w,
\end{equation}
where
\begin{align*}
  \rho(z,w)
  &= \exp\biggl(
       - \sigma^2 z\,\biggl(\frac{w}{\pi\lambda}\biggr)^{\beta/2}\!
     - w\,\biggl(
            1 + \frac{2 z^{2/\beta}}{\beta}
                \int_{1/z}^\infty\frac{v^{2/\beta-1}}{1+v}\,\dd v
          \biggr)
     \biggr).
\end{align*}  
\end{proposition}

\begin{IEEEproof}
For $\bm{u}\in\R^2$, let $B(\bm{u})$ denote the index of the nearest
point of $\Phi = \sum_{i\in\N}\delta_{X_i}$ to the location~$\bm{u}$;
that is, $\|X_{B(\bm{u})} - \bm{u}\| < \|X_i - \bm{u}\|$ for
$i\in\N\setminus\{B(\bm{u})\}$.
Suppose that the typical UE is located at $S(t) = \bm{u}$ at
slot~$t\in\N_0$.
Since $\sum_{i=1}^\infty\delta_{X_i-\bm{u}}$ is
equal in distribution to $\Phi$ due to the stationarity and $H_{i,t}$,
$i\in\N$, $t\in\N_0$, are i.i.d., we have from~\eqref{eq:SINR}
with~\eqref{eq:Interference} that $\SINR_{\bm{u},B(\bm{u})}(t)$ is
equal in distribution to $\SINR_{\bm{0},B(\bm{0})}(0)$ for any
$\bm{u}\in\R^2$.
Thus, since $\{S(t)\}_{t\in\N_0}$ is independent of $\Phi$ and
$\{H_{i,t}\}_{i\in\N,t\in\N_0}$, the definition of the expected
downlink data rate in \eqref{eq:T} leads to
\begin{align*}
  \mathcal{T}^{(0)}
  &= \lim_{m\to\infty}
       \frac{1}{m}\sum_{t=0}^{m-1}
         \Exp[\xi_{S(t),B(S(t))}(t)]
  \\
  &= \Exp[\xi_{\bm{0},B(\bm{0})}(0)],
\end{align*}
which implies that the expected downlink data rate in Scenario~0 is
equal to that for a static UE.
Hence, the existing result of the expected downlink data rate for a
static UE gives \eqref{eq:DataRate0} (see, e.g.,
\cite[Theorem~3]{AndrBaccGant11}).
\end{IEEEproof}

\begin{remark}
Proposition~\ref{prp:DataRate0} implies that, if a moving UE is always
connected to its nearest BS, the expected downlink data rate is
identical to that for a static UE.
Note that this fact is derived under the condition that $\Phi$ is
stationary, $H_{i,t}$, $i\in\N$, $t\in\N_0$, are i.i.d., and
$\{S(t)\}_{t\in\N_0}$ is independent of $\Phi$ and
$\{H_{i,t}\}_{i\in\N,t\in\N_0}$.
In other words, this holds true even when the locations of BSs are
according to a general stationary point process and the fading effects
independently follow a general identical distribution.
A similar discussion is found in \cite[Remark~2]{ChatBlasAltm19}.
\end{remark}

\subsubsection{Expected Downlink Data Rate in Scenario 1}
In Scenario~1, the typical UE is not always connected to its nearest
BS but remains connected to the BS that is the nearest at the
beginning of each cycle of the skipping period.

\begin{theorem}\label{thm:DataRate1}
For the cellular network model described in Sec.~\ref{sec:model},
the expected downlink data rate in Scenario~1 with $s$~slots of the
skipping period is given by
\begin{equation}\label{eq:DataRate1}
  \mathcal{T}^{(1)}
  = \frac{1}{s}\sum_{t=0}^{s-1}
      \int_0^\infty
        \tau(t v)\,
      \dd F_V(v),
\end{equation}
where $F_V$ denotes the distribution function of the moving
speed~$V_1=\|Y_1\|/s$ of the typical UE in a cycle of $s$~slots and
\begin{equation}\label{eq:tau1}
  \tau(u)
  = \int_0^\infty
      \frac{1}{z}\,
      \exp\bigl(-\sigma^2 z - \pi\lambda\,K_\beta\,z^{2/\beta}\bigr)\,
      \bigl(\mu(z,u) - 1\bigr)\,\dd z,
\end{equation}
with
\begin{align}
  K_\beta
  &= \frac{2\pi}{\beta}\,\csc\frac{2\pi}{\beta},
  \label{eq:K_beta}\\
  \mu(z,u)
  &= 2\pi\lambda\int_0^\infty
       r \exp\bigl(
           - \lambda\,\bigl[
               \pi r^2
               - J(r,z,u)
             \bigr] 
         \bigr)\,
     \dd r,
  \label{eq:rho1}\\
  J(r,z,u)
  &= 2 z
     \int_0^{\pi}\!\!\!\int_0^r
        \frac{x}
             {z + {w_{x,u,\phi}}^\beta}\,
      \dd x\,\dd\phi,
  \label{eq:J1}
\end{align}
and $w_{x,u,\phi}=\sqrt{x^2+u^2 - 2 x u\cos\phi}$.
\end{theorem}

The proof of Theorem~\ref{thm:DataRate1} relies on the following
lemma.

\begin{lemma}\label{lem:tau1}
Suppose that a UE is located at $\bm{u}\in\R^2$ with $\|\bm{u}\|=u$ at
slot $t\in\{0,1,\ldots,s-1\}$ and is served by the BS at
$X_{B(\bm{0})}$, which is the nearest BS to the origin.
Then, the expected instantaneous downlink data rate for this UE
satisfies $\Exp[\xi_{\bm{u},B(\bm{0})}(t)] = \tau(u)$ given
in~\eqref{eq:tau1}.
\end{lemma}

\begin{IEEEproof}
See Appendix~\ref{app:Prf_Lemma_tau1}
\end{IEEEproof}

\begin{IEEEproof}[Proof of Theorem~\ref{thm:DataRate1}]
As in the proof of Proposition~\ref{prp:DataRate0}, let $B(\bm{u})$
denote the index of the nearest point of $\Phi =
\sum_{i\in\N}\delta_{X_i}$ to $\bm{u}\in\R^2$.
In Scenario~1, we see from~\eqref{eq:P_ns} and~\eqref{eq:P_t} that the
typical UE is connected to the BS at $X_{B(S(ns))}$ at slot~$ns+t$
for $n\in\N_0$ and $t\in\{0,1,\ldots,s-1\}$.
Thus, the expected downlink data rate in~\eqref{eq:T} is reduced to
\begin{align}
  \mathcal{T}^{(1)}
  &= \lim_{m\to\infty}\frac{1}{m}
       \sum_{n=0}^{\lfloor m/s\rfloor}\sum_{t=0}^{s-1}
         \Exp\bigl[\xi_{S(ns+t),B(S(ns))}(ns+t)\bigr]
  \nonumber\\
  &= \frac{1}{s}\sum_{t=0}^{s-1}
       \Exp[\xi_{S(t),B(\bm{0})}(t)],
\end{align}  
where the last equality follows from the distributional equivalence of
$\SINR_{S(ns+t),B(S(ns))}(ns+t)$ and $\SINR_{S(t),B(\bm{0})}(t)$ for
$t\in\{0,1,\ldots,s-1\}$, which follows because $\Phi$ is stationary
and isotropic, $H_{i,t}$, $i\in\N$, $t\in\N_0$, are i.i.d., and also
$Y_k$, $k\in\N$, in~\eqref{eq:P_ns} are i.i.d.\ and independent of
$\Phi$ and $\{H_{i,t}\}_{i\in\N, t\in\N_0}$.
Hence, we obtain \eqref{eq:DataRate1} since
$\Exp[\xi_{S(t),B(\bm{0})}(t)\mid S(t)=\bm{u}] = \tau(t v)$ when
$\|\bm{u}\| = t v$ by Lemma~\ref{lem:tau1} and $\|S(t)\| =
(t/s)\,\|Y_1\| =  t V_1$ for $t\in\{0,1,\ldots,s-1\}$
by~\eqref{eq:P_t}.
\end{IEEEproof}

The expressions~\eqref{eq:DataRate1}--\eqref{eq:J1} obtained in
Theorem~\ref{thm:DataRate1} are indeed numerically computable.
However, they include some nested integrals, which may annoy us with a
heavy computational load.
In the rest of this subsection, we discuss some ways of reducing the
computational load.

\subsubsection{Tips for Computational Load Reduction}

We here introduce some tips to reduce the load of computing
$\mathcal{T}^{(1)}$ in Theorem~\ref{thm:DataRate1} exactly or
approximately.
First, we find that a simple change of variables reduces the number of
nested integrals.

\begin{lemma}\label{lem:J1}
Function~$J$ in \eqref{eq:J1} is equal to the following.
\begin{equation}\label{eq:J11}
  J(r,z,u)
  = 2 z
    \int_0^{u+r}
      \frac{x}{z + x^\beta}\,
       C(x,r,u)\,      
     \dd x,
\end{equation}    
with
\[
  C(x,r,u)
  = \begin{cases}
      \pi, &\quad u=0,
      \\
      \displaystyle{%
        \arccos\Bigl(
          -1\vee\frac{x^2+u^2-r^2}{2x u}\wedge 1
        \Bigr),
      }&\quad u>0,
    \end{cases}
\]
where $a\vee b=\max\{a,b\}$ and $a\wedge b=\min\{a,b\}$ for $a,
b\in\R$.
\end{lemma}  

\begin{IEEEproof}
See Appendix~\ref{app:Prf_Lem_J1}
\end{IEEEproof}

\begin{table}[!t]
%\renewcommand{\arraystretch}{1.3}  
%\caption{Comparison of the computation time of $\mathcal{T}^{(1)}$
%  using \eqref{eq:J1} and \eqref{eq:J11}.
%  The parameter values are fixed at $\lambda=10$~({\upshape
%    $\text{units}/\text{km}^2$}), $\beta=3$ and $\sigma^2=5$, and the
%  moving speed of the typical UE is constant at $v=0.005$~({\upshape
%    km/sec}).}%
\caption{Comparison of the computation time of $\tau(u)$ using
  \eqref{eq:J1} and \eqref{eq:J11}. 
  The parameter values are fixed at $\lambda=10$~({\upshape
    $\text{units}/\text{km}^2$}), $\beta=3$ and $\sigma^2=25$.}%
\label{table:cmprsn_J}
\centering
\begin{tabular}{ccccccc}
  \hline
%  $s$~($\times1000$ msec) & 5 & 10 & 20 & 50 & 100 \\
%  \hline
%  Use of Eq.~\eqref{eq:J1}~(sec)  & 3108  & 11026 & 38350 & 171248 & 423740 \\
%  Use of Eq.~\eqref{eq:J11}~(sec) & 2394  & 6534  & 14870 & 38751  &  76353 \\
%  Reduction rate~(\%)             & 22.97 & 40.74 & 61.23 & 77.37  & 81.98 \\
%
  $u$~(km) & 0.1 & 0.2 & 0.3 & 0.4 & 0.5 & 0.6\\
  \hline
  Use of eq.~\eqref{eq:J1}~(sec)
  & 1344 & 973  & 10221 & 7286 & 4754 & 1474 \\
  Use of eq.~\eqref{eq:J11}~(sec)
  & 695  & 1118 & 1196  & 632  & 974  & 401  \\
%  Reduction rate~(\%)
%  & 48.29 & $-14.90$ & 88.30 &  91.33  & 79.51 & 72.80\\
  \hline
\end{tabular}  
\end{table}

We can observe through experiments that \eqref{eq:J11} reduces the
computation time of the expected instantaneous downlink data
rate~$\tau(u)$ by about 60\% on average compared to the use of
\eqref{eq:J1} (see Table~\ref{table:cmprsn_J}).
Next, we give a simple lower bound for $\tau$ in~\eqref{eq:tau1} under
the interference-limited (noise-free) assumption.

\begin{corollary}\label{cor:tau1_lower}
Suppose that $\sigma^2=0$.
Then, $\tau$ in \eqref{eq:tau1} is bounded below as follows.
\begin{equation}\label{eq:tau1_lower}
  \tau(u)
  \ge \frac{\beta}{2}
      \int_0^\infty
        \frac{z^{\beta/2-1}}
             {(1+z)({K_\beta}^{\beta/2}+z^{\beta/2})}\,
        \exp\biggl(
          -\pi\lambda\,u^2\,\frac{z}{1+z}
        \biggr)\,
      \dd z,
  \quad u\ge0,
\end{equation} 
with $K_\beta$ given in~\eqref{eq:K_beta}.
\end{corollary}

\begin{IEEEproof}
See Appendix~\ref{app:Prf_Cor_tau1}.
\end{IEEEproof}

\begin{remark}
As we can see in the proof, the lower bound in
Corollary~\ref{cor:tau1_lower} is obtained by relaxing the condition
that there must not be other BSs closer than the nearest BS to the
origin.
Similar bounds/approximates are often found in the literature (see,
e.g., \cite{NguyBaccKofm07}).
\end{remark}

\begin{figure}[!t]
  \centering
  \includegraphics[width=.5\linewidth]{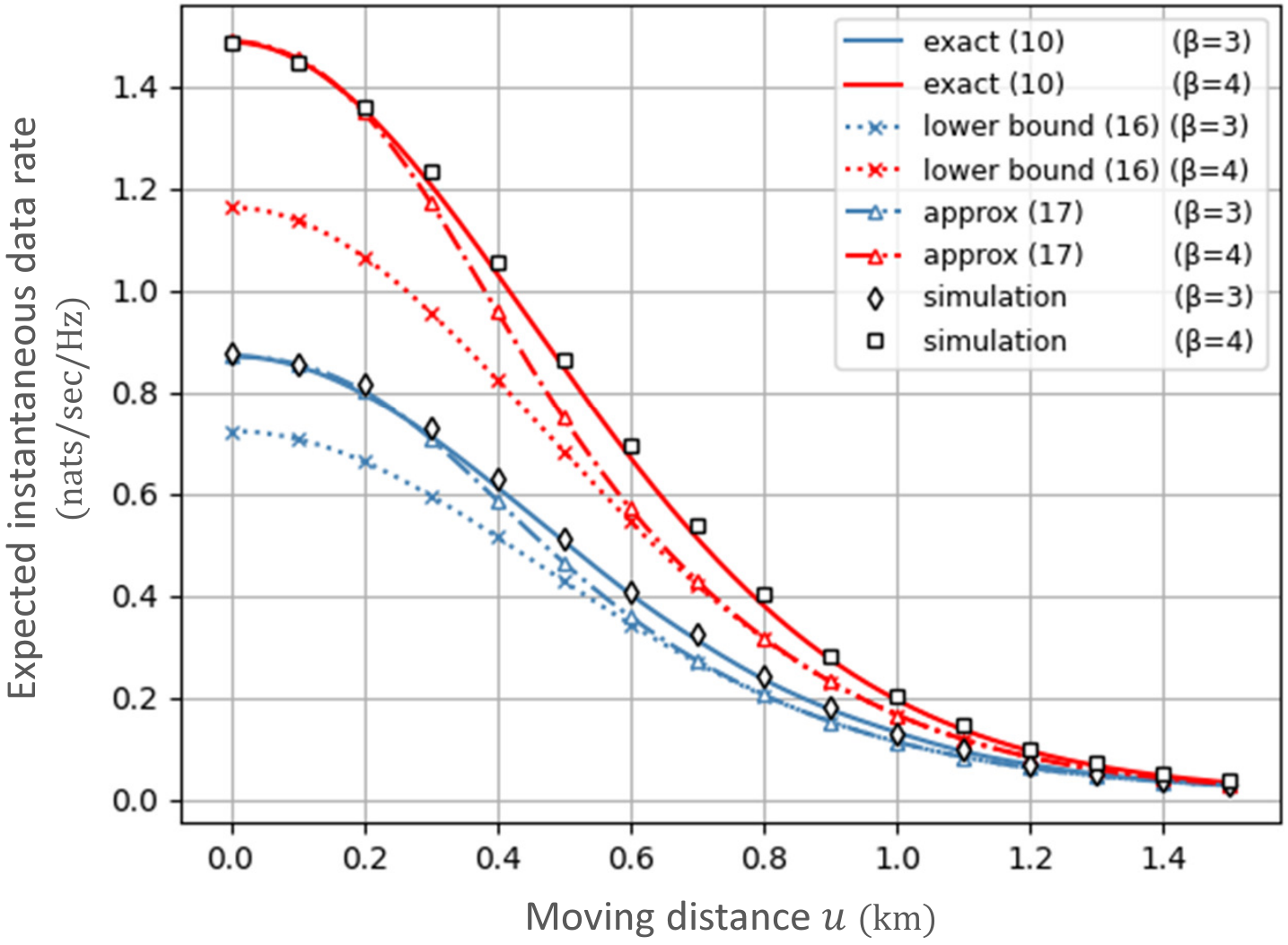}
  \caption{Numerical comparison of $\tau$ in~\eqref{eq:tau1} with the
    lower bound~\eqref{eq:tau1_lower}, the
    approximation~\eqref{eq:tau1_approx}, and the values from Monte
    Carlo simulation.
    The BS intensity is fixed at $\lambda=1$
    ($\text{units}/\text{km}^2$) and two patterns of $\beta=3$ and
    $\beta=4$ are exhibited.}%
  \label{fig:cmprsn_tau}
\end{figure}

The lower bound obtained in Corollary~\ref{cor:tau1_lower} is indeed
of a simple form (including a single integral), but as seen in
Fig.~\ref{fig:cmprsn_tau}, it causes non-negligible gaps from the
exact values, in particular when the moving distance $u$ is small,
whereas the gaps decrease as $u$ increases.
On the other hand, we know that $\tau(0)=\mathcal{T}^{(0)}$ since the
expected downlink data rate in Scenario~0 is equal to that for a
static UE.
Hence, we can approximate $\tau$ in \eqref{eq:tau1} by interpolating
between $\mathcal{T}^{(0)}$ and the lower bound in
Corollary~\ref{cor:tau1_lower} as follows.
Suppose $\sigma^2=0$ as in Corollary~\ref{cor:tau1_lower} and let
$\Tilde{\tau}$ denote the lower bound of $\tau$ given on the
right-hand side of \eqref{eq:tau1_lower}.
Then, $\tau$ in \eqref{eq:tau1} is approximated as
\begin{equation}\label{eq:tau1_approx}
  \tau(u)
  \approx \epsilon(u)\,\mathcal{T}^{(0)}
          + (1-\epsilon(u))\,\Tilde{\tau}(u),
  \quad u\ge0,
\end{equation}
where a function~$\epsilon$:~$[0,\infty)\to[0,1]$ is smooth and
decreasing, and satisfies $\epsilon(0)=1$ and $\epsilon(u)\to0$ as
$u\to\infty$; that is, it is chosen in such a way that the right-hand
side of \eqref{eq:tau1_approx} is close to $\mathcal{T}^{(0)}$ when
$u$ is small, and it approaches $\Tilde{\tau}(u)$ as $u$ becomes
larger.
Figure~\ref{fig:cmprsn_tau} compares the numerical results of $\tau$
in \eqref{eq:tau1} with its lower bound in~\eqref{eq:tau1_lower} and
the approximation in~\eqref{eq:tau1_approx}, as well as with the
values from Monte Carlo simulation, with respect to the moving
distance~$u$.
In the approximation~\eqref{eq:tau1_approx}, the function~$\epsilon$
is set as $\epsilon(u)= e^{-10\,u^2}$, $u\ge0$.
The simulation results are computed as the mean of 10,000 independent
samples.
As stated above, the values of the lower bound have some gaps from the
exact values when $u$ is small, whereas these gaps decrease as $u$
increases.
This implies that the condition that other BSs never exist closer than
the nearest BS is nonnegligible when the typical UE is close to the
origin, but it is diminishing as the UE moves away from the origin.
On the other hand, the approximation~\eqref{eq:tau1_approx} shows good
agreement with the exact values as expected.
However, we should note that such agreement depends on a choice of the
function~$\epsilon$.
An exponential function~$\epsilon(u)=e^{-a u^b}$ as above seems an
plausible choice as one with the desired properties (that is, smooth
and decreasing with $\epsilon(0)=1$ and
$\lim_{u\to\infty}\epsilon(u)=0$), and statistical fitting of $a$ and
$b$ would lead to better results.

\subsection{Handover Rate Analysis}

We now proceed to the analysis of the HO rate.
Similar to the proof of Theorem~\ref{thm:DataRate1}, the HO rate
in~\eqref{eq:H} is reduced to
\begin{align}\label{eq:HO_Rate}
  \mathcal{H}
  &= \lim_{m\to\infty}
       \frac{1}{m}\sum_{n=0}^{\lfloor m/s\rfloor}
         \Exp[\zeta(ns,(n+1)s)]
  \nonumber\\    
  &= \frac{\Exp[\zeta(0,s)]}{s},
\end{align}
where the last equality follows since $\Phi$ is stationary and
isotropic, and $Y_k$, $k\in\N$, in~\eqref{eq:P_ns} are i.i.d.\ and
independent of $\Phi$.
By \eqref{eq:HO_Rate}, it is enough to consider the expected number of
HOs in a cycle of $s$~slots, during which the typical UE moves along a
straight line segment, and we can use the existing results in both
Scenarios~0 and 1.

\subsubsection{HO Rate in Scenario~0}

The HO rate in the scenario without any HO skipping has so far been
studied in the literature.
The following is a direct consequence of it.

\begin{proposition}\label{prp:HORate0}
For the cellular network model described in Sec.~\ref{sec:model},
the HO rate in Scenario~0 is given by
\begin{equation}\label{eq:HORate0}
  \mathcal{H}^{(0)}
  = \frac{4\sqrt{\lambda}\,\Bar{v}}{\pi},
\end{equation}    
where $\Bar{v}$ denotes the average moving speed of the typical UE in
a cycle of $s$~slots; that is, $\Bar{v} = \Exp[V_1]$ with
$V_1=\|Y_1\|/s$.
\end{proposition}

\begin{IEEEproof}
Given $L_1=\|Y_1\|=\ell$, the conditionally expected number of
HOs~$\Exp[\zeta^{(0)}(0,s)\mid L_1=\ell]$ in a cycle is equal to the
expected number of intersections of a line segment of length~$\ell$
with the boundaries of the Poisson-Voronoi cells, and is well-known as
$\Exp[\zeta^{(0)}(0,s)\mid L_1=\ell] = 4\sqrt{\lambda}\,\ell/\pi$
(see, e.g., \cite{Moll89,BaccZuye99,LinGantFlemAndr13}).
Applying this to \eqref{eq:HO_Rate} with $L_1=s V_1$ derives
\eqref{eq:HORate0} by taking the expectation.
\end{IEEEproof}

\subsubsection{HO Rate in Scenario~1}

The HO rate in a similar scenario to our Scenario~1 is studied
in~\cite{SadrAdve15}, which helps us to show the following.

\begin{proposition}\label{prp:HORate1}
For the cellular network model described in Sec.~\ref{sec:model},
the HO rate in Scenario~1 with $s$~slots of the skipping period is
given by
\begin{equation}\label{eq:HORate1}
  \mathcal{H}^{(1)}
  = \frac{1}{s}\biggl[
      1 - 2\lambda
          \int_0^\infty\!\!\!\int_0^\pi\!\!\!\int_0^\infty
            r\,e^{- \lambda\,\eta(r, s v, \phi)}\,
          \dd r\,\dd\phi\,\dd F_V(v)
    \biggr],
\end{equation}    
where
\begin{equation}\label{eq:eta}
  \eta(r, \ell, \phi)
  = {w_{r,\ell,\phi}}^2\,
    \arccos\biggl(
      \frac{r\cos\phi-\ell}{w_{r,\ell,\phi}}
    \biggr)
    + r^2(\pi-\phi) + r\ell\sin\phi,
\end{equation}
with $w_{r,\ell,\phi}=\sqrt{r^2+\ell^2-2 r\ell\cos\phi}$, and $F_V$ is
(as in \eqref{eq:DataRate1}) the distribution function of the moving
speed~$V_1=\|Y_1\|/s$ of the typical UE in a cycle of $s$~slots.
\end{proposition}

\begin{IEEEproof}
By the isotropy of $\Phi=\sum_{i\in\N}\delta_{X_i}$, we can assume
without loss of generality that the typical UE moves in the positive
direction along the horizontal axis during a cycle of $s$~slots.
Suppose that the typical UE initially connected to the BS at
$X_{B(\bm{0})} = \bm{x} = (r,\phi)$ in the polar coordinates and moves
to $Y_1 = \bm{y} =(\ell,0)$ in $s$~slots.
Let $b_{\bm{x}}(r)$ denote the disk centered at $\bm{x}\in\R^2$ with
radius~$r>0$.
Since there are no BSs in $b_{\bm{0}}(r)$ and the distance to the
initial BS at $\bm{x}$ from the location~$\bm{y}$ is equal
to $w_{r,\ell,\phi} = \sqrt{r^2+\ell^2-2r\ell\cos\phi}$, the typical
UE executes an HO at the end of the cycle if and only if there is at
least one BS in the area~$b_{\bm{y}}(w_{r,\ell,\phi})\setminus
b_{\bm{0}}(r)$.
Hence, similar discussion to \cite{SadrAdve15} gives
\begin{align}\label{eq:prf_prp_HO1}
  &\Exp\bigl[
     \zeta^{(1)}(0,s)\mid X_{B(\bm{0})}=(r,\phi), Y_1=(\ell,0)
   \bigr]
  \nonumber\\
  &= 1 - e^{-\lambda\,|b_{\bm{y}}(w_{r,\ell,\phi})\setminus b_{\bm{0}}(r)|}
  \nonumber\\
  &= 1 - \exp\biggl(
           -\lambda\,\biggl[
             {w_{r,\ell,\phi}}^2\,
             \arccos\biggl(
               \frac{r\cos\phi-\ell}{w_{r,\ell,\phi}}
             \biggr)
             - r^2\phi + r\ell\sin\phi
            \biggr]
         \biggr),
\end{align}
where $|A|$ denotes the Lebesgue measure of $A\in\B(\R^2)$.
This can be unconditioned by integrating with respect to the density
$f_0(r)\,\dd r = 2\pi\lambda r\,e^{-\pi\lambda r^2}\,\dd r$ of
$\|X_{B(\bm{0})}\|$ over $[0,\infty)$, $\dd\phi/\pi$ over $[0,\pi)$,
    and $\dd F_V(v)$ over $[0,\infty)$ with $v=\ell/s$.
Finally, plugging it into \eqref{eq:HO_Rate}, we have
\eqref{eq:HORate1}.
\end{IEEEproof}

%Note that, in Scenario~1, $\Exp[\zeta^{(1)}(0,s)] =
%s\,\mathcal{H}^{(1)}$ represents the probability that the typical UE
%executes an HO at the end of the skipping period.

\subsection{Numerical Evaluation of Performance Metrics}

\begin{figure}[!t]
  \centering
  \subfloat[The expected downlink data rate.]{%
    \includegraphics[width=.5\linewidth]{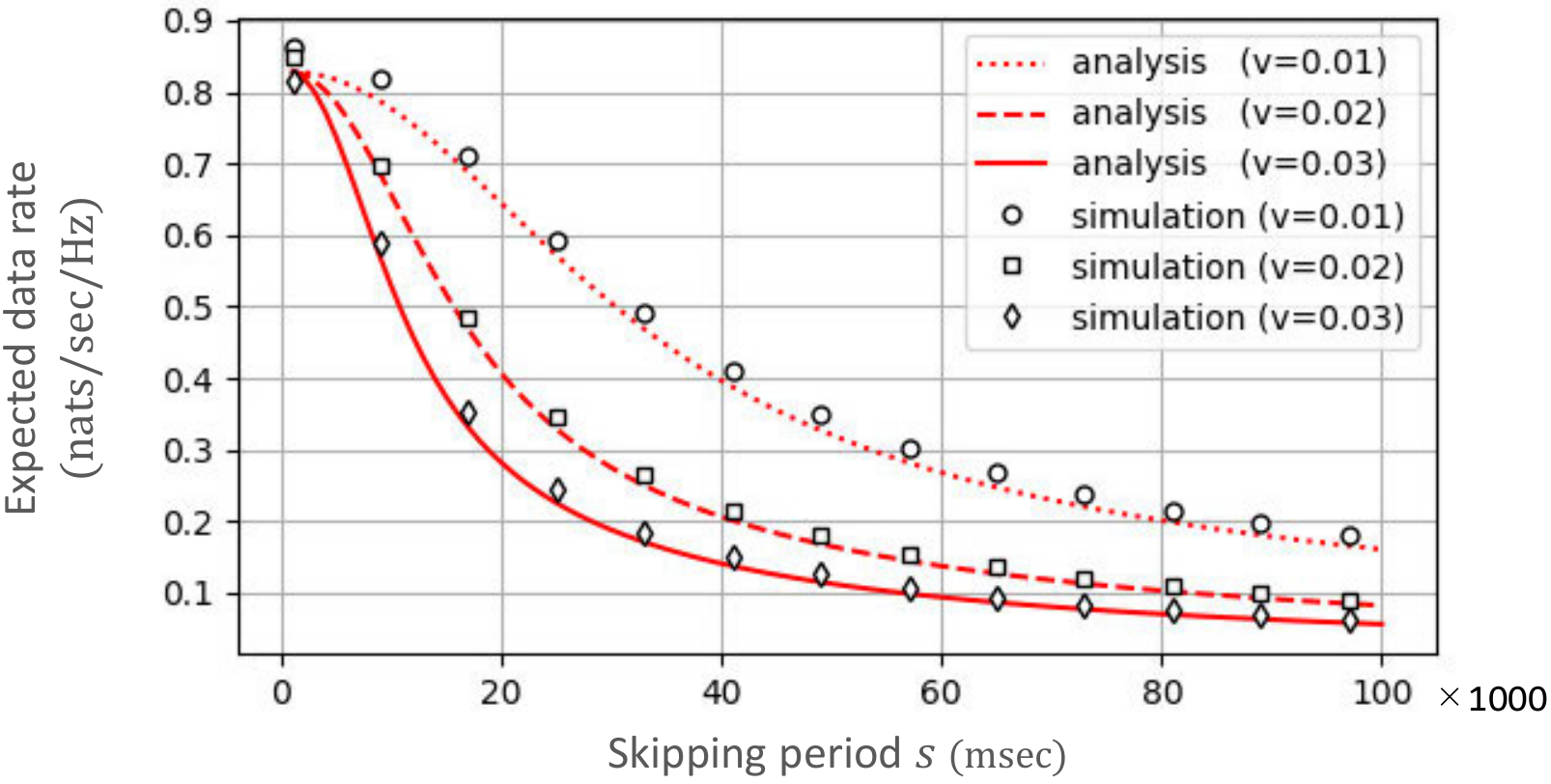}%
    \label{subfig:DataRate}%
  }%
  \\
  \subfloat[The HO rate.]{%
    \includegraphics[width=.5\linewidth]{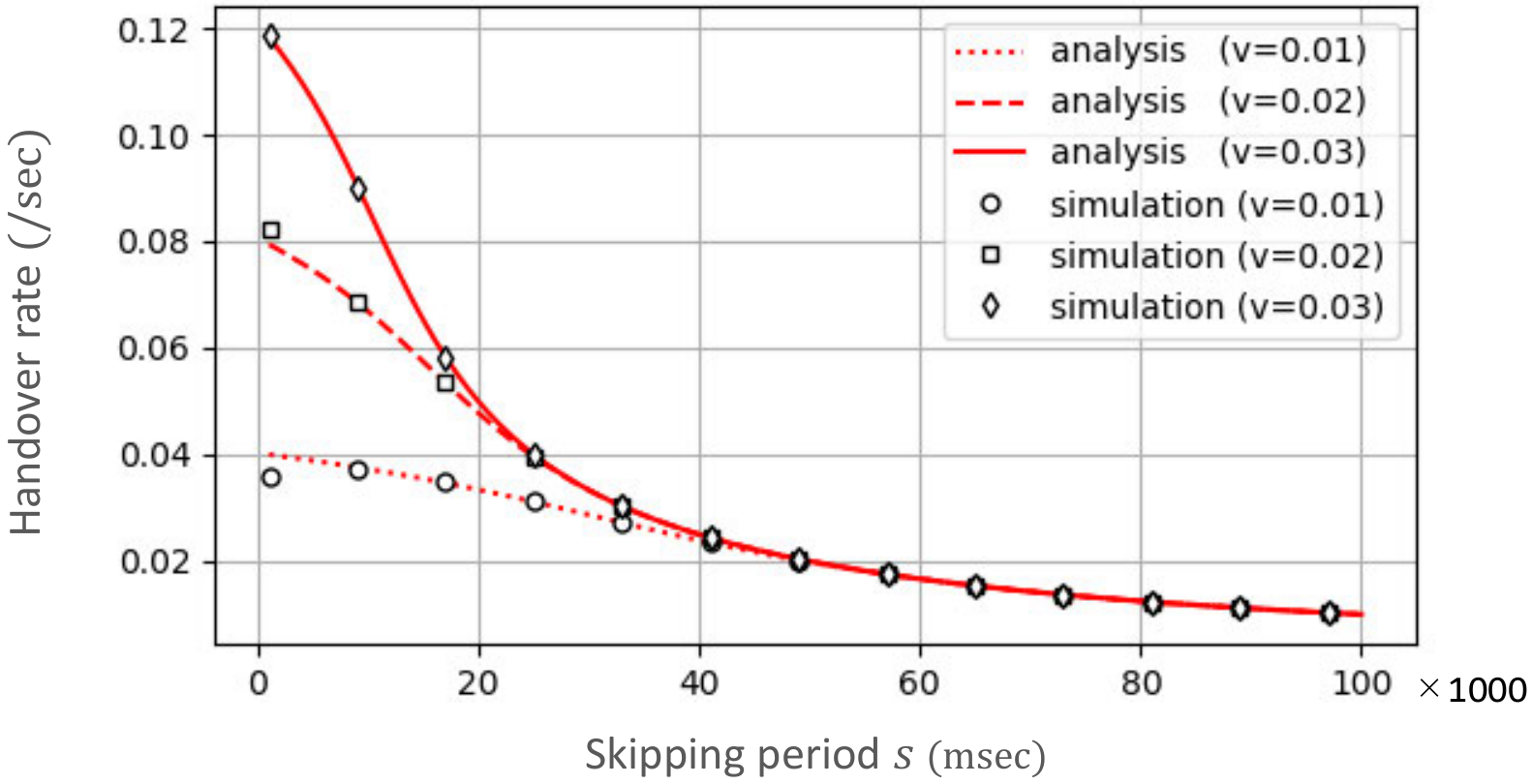}%
    \label{subfig:HORate}%
  }%
  \caption{The performance metrics as functions of the length~$s$ of
    the skipping period for three patterns of the moving speed
    $v=0.01$, $0.02$, and $0.03$~(km/sec).}
  \label{fig:DataRate&HORate}
\end{figure}

We here numerically evaluate the expected downlink data
rate and the HO rate in the periodic HO skipping scheme, which are
respectively obtained in Theorem~\ref{thm:DataRate1} (with
Lemma~\ref{lem:J1}) and Proposition~\ref{prp:HORate1}.
Throughout the experiments, we set as $1\text{slot}=1\text{msec}$, the
intensity of the BSs, the path-loss exponent, and the noise power are
fixed at $\lambda=10$~(units/km$^2$), $\beta=3$, and $\sigma^2=25$,
respectively, and the moving speed $V_1=\|L_1\|/s$ of the typical UE
is given as a constant.
Figure~\ref{fig:DataRate&HORate} shows the curves of
$\mathcal{T}^{(1)}$ and $\mathcal{H}^{(1)}$ with respect to the
length~$s$ of the skipping period.
For comparison, the values from Monte Carlo simulation are also
plotted as the means of 1,000 independent samples of
$\sum_{t=0}^{m-1}\xi^{(1)}(t)/m$ and $\zeta^{(1)}(0,m)/m$,
respectively with $m=1000$ (cf.~\eqref{eq:T} and \eqref{eq:H}).
%of $\mathcal{T}^{(1)}$ and $\mathcal{H}^{(1)}$ 
We find that the analytical results match well with the simulation
results.
Moreover, we can confirm the trade-off relation between the HO rate
and the expected downlink data rate; that is, both are decreasing in
the length of the skipping period.
We further explore this trade-off in the next subsection.

\subsection{Utility Metric}

To discuss the trade-off between the expected downlink data
rate~$\mathcal{T}$ and the HO rate~$\mathcal{H}$, we introduce a
utility metric~$\mathcal{U}$ as
\begin{equation}\label{eq:U}
  \mathcal{U}
  = \mathcal{T} - c\,\mathcal{H},
\end{equation}
where a utility constant~$c>0$ is suitably chosen so as to convert the
negative impact of HOs into the loss of the downlink data rate.
Note that similar metrics are found in the literature
(cf.~\cite{BaoLian15,BaoLian16,ArshElSaSoroAlNaAlou17,TabaSaleHoss19})
and are often referred to as the user throughput accounting for the
loss due to HOs.
However, we do not use the term ``throughput'' in this paper because
$\mathcal{U}$ in \eqref{eq:U} can take negative values (see, e.g.,
Figs.~\ref{fig:Compare_Scenarios}
and~\ref{fig:Utility-vs-SkippingTime} below).

Figure~\ref{fig:Compare_Scenarios} compares the utility
metrics~$\mathcal{U}^{(0)}$ and $\mathcal{U}^{(1)}$ respectively for
Scenarios~0 and 1 with respect to the average speed~$\Bar{v} =
\Exp[V_1]$ of the typical UE.
In the computation of $\mathcal{U}^{(1)}$, $\tau$ in
\eqref{eq:DataRate1} is replaced by its approximation
\eqref{eq:tau1_approx} with the adjustment function~$\epsilon(u) =
e^{-10\,u^2}$.
Four different distributions of the moving speed are experimented with
the common average~$\Bar{v}$; that is, exponential, second-order
Erlang, second-order hyper-exponential, and deterministic ones, where
in the hyper-exponential distribution, two exponential distributions
with means~$\Bar{v}/2$ and $3\Bar{v}/2$ are mixed with equal
probability.
The other parameters are fixed as $\lambda=1$~(units/km$^2$),
$\beta=3$, $\sigma^2=0$, $s=50,000$~(msec), and $c=10$~(nats/Hz).
Note that only one line is exhibited for $\mathcal{U}^{(0)}$ since it
depends on the distribution of the moving speed only through its
average (as confirmed from \eqref{eq:DataRate0} and
\eqref{eq:HORate0}).
From Fig.~\ref{fig:Compare_Scenarios}, we observe that Scenario~0
shows better performance when the average moving speed is small
(roughly $\Bar{v}\le 0.05$~(km/sec)), whereas Scenario~1 becomes
better as the UE moves faster.
This is thought to be because $\mathcal{H}^{(0)}$ is linearly
increasing in $\Bar{v}$ (see \eqref{eq:HORate0}), whereas $\tau(u)$ in
\eqref{eq:tau1} is slowly decreasing in $u$ (see
Fig.~\ref{fig:cmprsn_tau}).
Moreover, we find an interesting property from the experiment that the
distribution of the moving speed has an impact on the utility metric
in Scenario~1; that is, the utility metric takes larger values as the
distribution of the moving speed is larger in variation.
Exploration of this property will be left for future work.

\begin{figure}[!t]
  \centering
  \includegraphics[width=.5\linewidth]{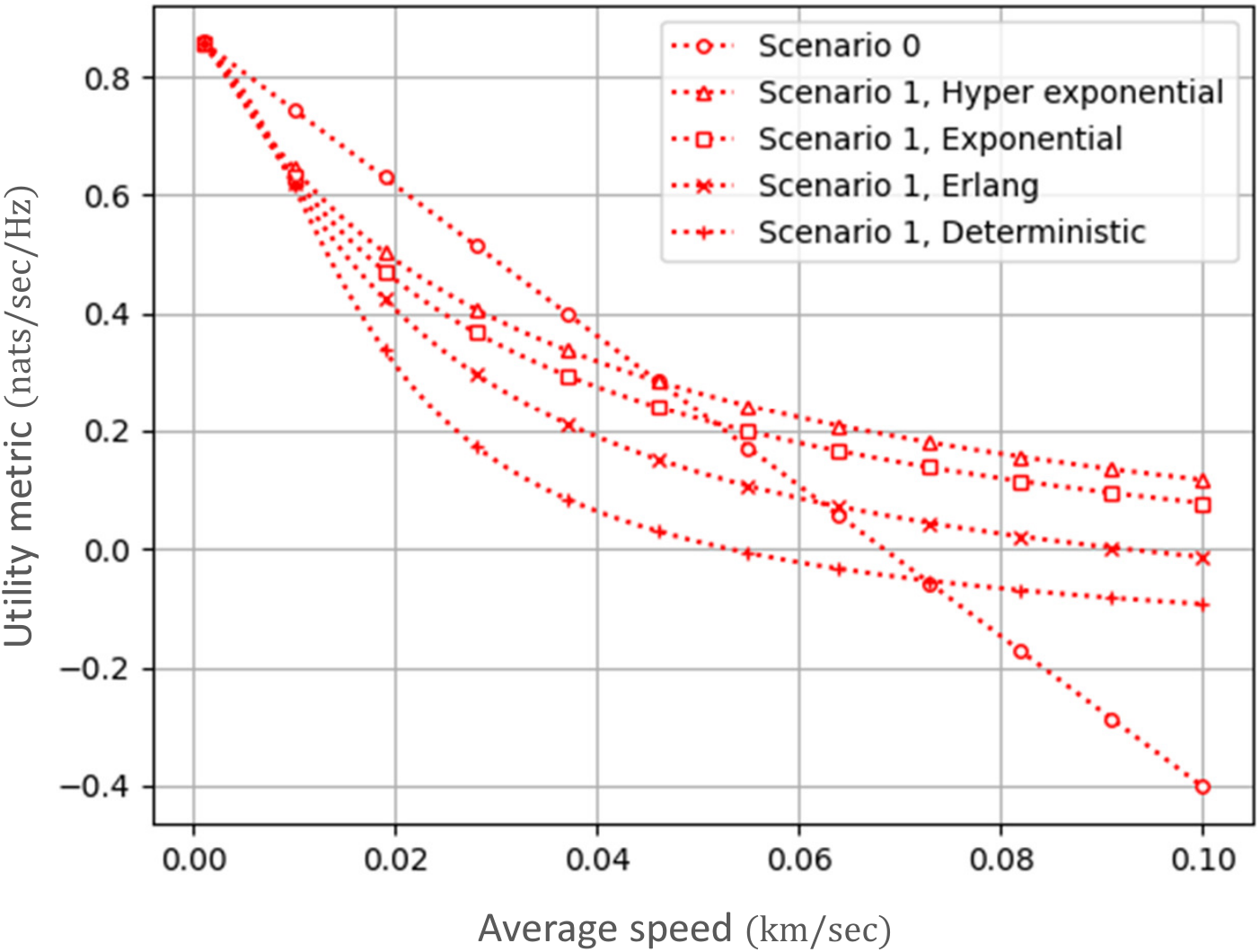}
  \caption{The values of $\mathcal{U}^{(0)}$ and $\mathcal{U}^{(1)}$
    as functions of the average speed~$\Bar{v}$ of the moving UE with
    several distributions.}
\label{fig:Compare_Scenarios}
\end{figure}

\section{Optimal Skipping Period}\label{sec:s_opt}

As stated in Sec.~\ref{subsec:HOSkipping}, the choice of the length of
the skipping period is vital for our proposed scheme.
In this section, we discuss how to decide the length~$s$ of the
skipping period on the basis of the analysis in the preceding section.
We here consider only Scenario~1---the periodic HO skipping
scenario---so that we omit the superscript~``(1)'' and just write
$\mathcal{T}$, $\mathcal{H}$ and $\mathcal{U}$.

\subsection{Approximate Derivation of Optimal Skipping Period}

\begin{figure}[!t]
  \centering
  \includegraphics[width=.5\linewidth]{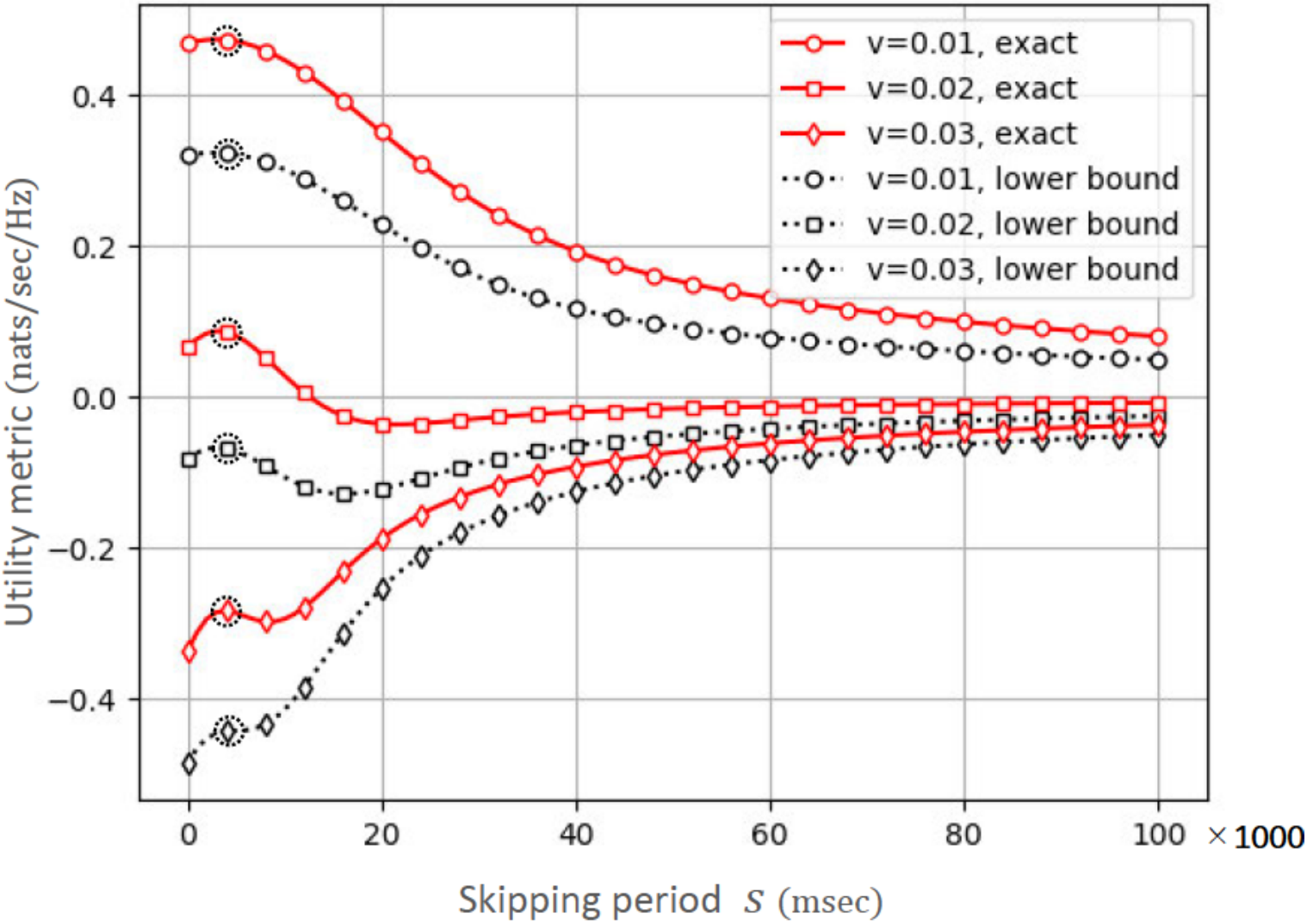}
  \caption{The values of utility metric and its lower bound as
    functions of the skipping
    period~$s$.}%
  \label{fig:Utility-vs-SkippingTime}
\end{figure}

Let us see Fig.~\ref{fig:Utility-vs-SkippingTime}, where the
numerical results of the utility
metric~$\mathcal{U}$ in \eqref{eq:U} and its lower bound, obtained by
replacing $\tau$ in \eqref{eq:DataRate1} with the right-hand side of
\eqref{eq:tau1_lower}, are plotted as functions of the length~$s$ of
the skipping period (in $1\text{slot}=1\text{msec}$) for three
different values of constant moving speed of the typical UE.
In this experiment, the parameters are set as
$\lambda=10$~(units/km$^2$), $\beta=3$, $\sigma^2=0$, and
$c=10$~(nats/Hz).
From the figure, we find that the utility metric has a local
maximum (at around $s=4,000$~(msec)) and this local maximum looks
global when the moving speed is small.
We refer to the length~$s$ of the skipping period which gives the
local maximum of the utility metric as the \textit{optimal skipping
  period}.
Furthermore, we should notice that the values of the optimal skipping
period are almost the same as those locally maximizing the lower bound
of $\mathcal{U}$ and are hardly affected by the difference in the
moving speed.
This suggests that we can approximately obtain the optimal skipping
period by finding the length~$s$ which locally maximizes the lower
bound of $\mathcal{U}$ for a certain moving speed of the typical UE.
From this observation, we have the following.

\begin{theorem}\label{thm:s_opt}
For the cellular network model described in Sec.~\ref{sec:model},
we suppose that the typical UE adopts the periodic HO skipping and
moves at certain constant speed.
Then, an approximation of the optimal skipping period is obtained as
the closest integer of $s^*$ given by
\begin{equation}\label{eq:s_opt}
  s^*
  = \Bigl(\frac{15}{\pi^2} -1\Bigr)
    \frac{c}{2\beta}
    \biggl(
      \int_0^\infty
        \frac{z^{\beta/2}}
             {(1+z)^2({K_\beta}^{\beta/2} + z^{\beta/2})}\,
      \dd z
    \biggr)^{-1},
\end{equation}
where $K_\beta$ is given in~\eqref{eq:K_beta}.
\end{theorem}

\begin{IEEEproof}
The proof approximately derives the length~$s$ of the skipping period
which maximizes the lower bound of the utility metric for sufficiently
small moving speed of the typical UE.
The details are given in Appendix~\ref{app:Prf_Thm_s_opt}.
\end{IEEEproof}

\begin{remark}
Theorem~\ref{thm:s_opt} is shown under the condition that the moving
speed~$v$ of the typical UE is sufficiently small (so that the terms
of $o(v^2)$ are negligible).
Indeed, as seen in Fig.~\ref{fig:Utility-vs-SkippingTime}, the optimal
skipping period is hardly affected by the difference in the moving
speed within the range of the experiment.
A further advantage of $s^*$ in \eqref{eq:s_opt} is that it is
determined only by the path-loss exponent~$\beta$ and the utility
constant~$c$ introduced in \eqref{eq:U}, but does not depend on the BS
intensity as well as the moving speed.
In other words, we can use $s^*$ in \eqref{eq:s_opt} regardless of the
moving speed and the BS intensity.
%The validity of the use of $s^*$ is confirmed
Some properties of $s^*$ are revealed through numerical experiments in
the next subsection.
\end{remark}

\subsection{Numerical Evaluation of $s^*$}

\begin{figure}[!t]
  \centering
  \subfloat[$s^*$ as a function of $\beta$.]{%
    \includegraphics[width=.5\linewidth]{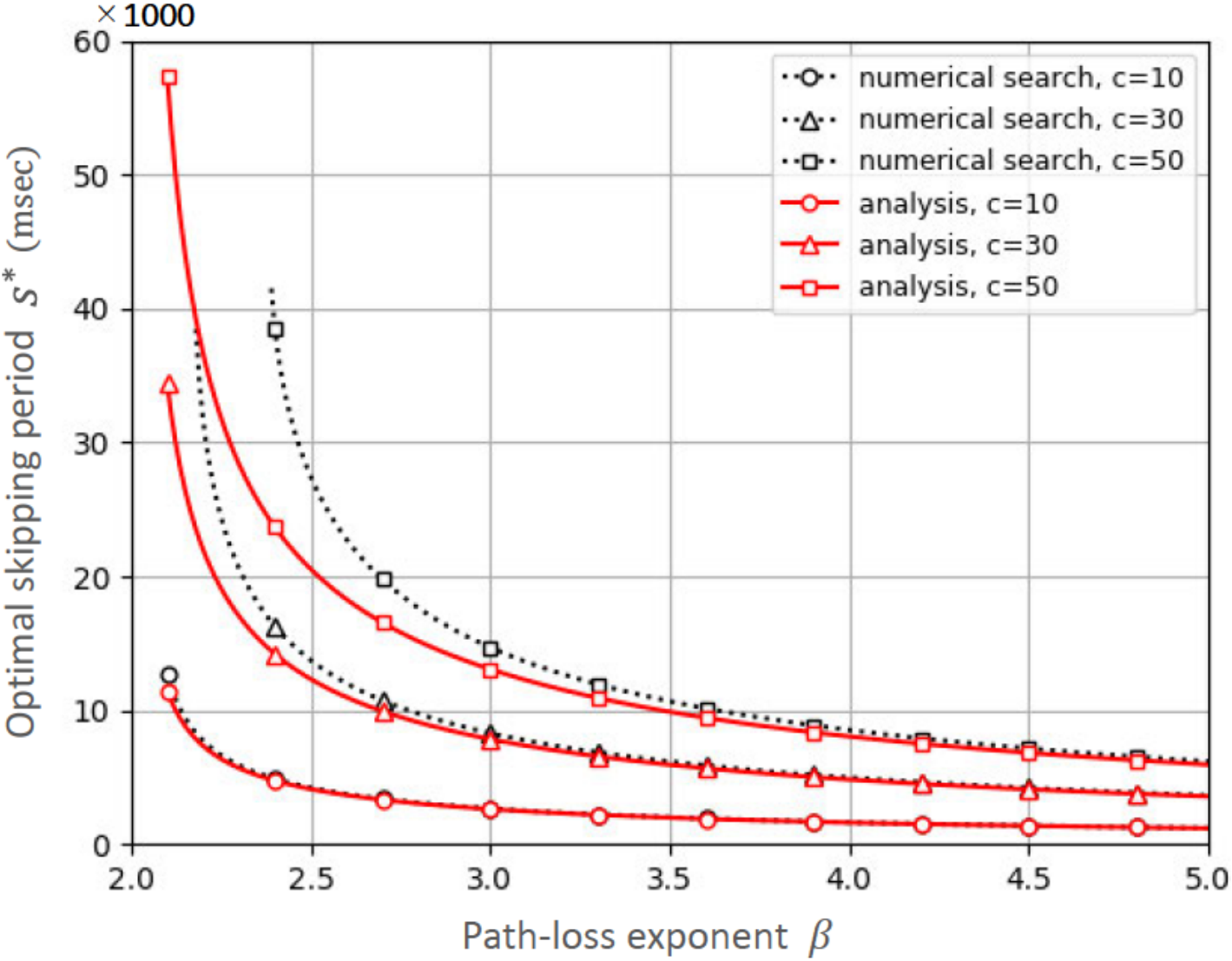}%
    \label{subfig:vsbeta}%
  }%
  \\
  \subfloat[$s^*$ as a function of $c$.]{%
    \includegraphics[width=.5\linewidth]{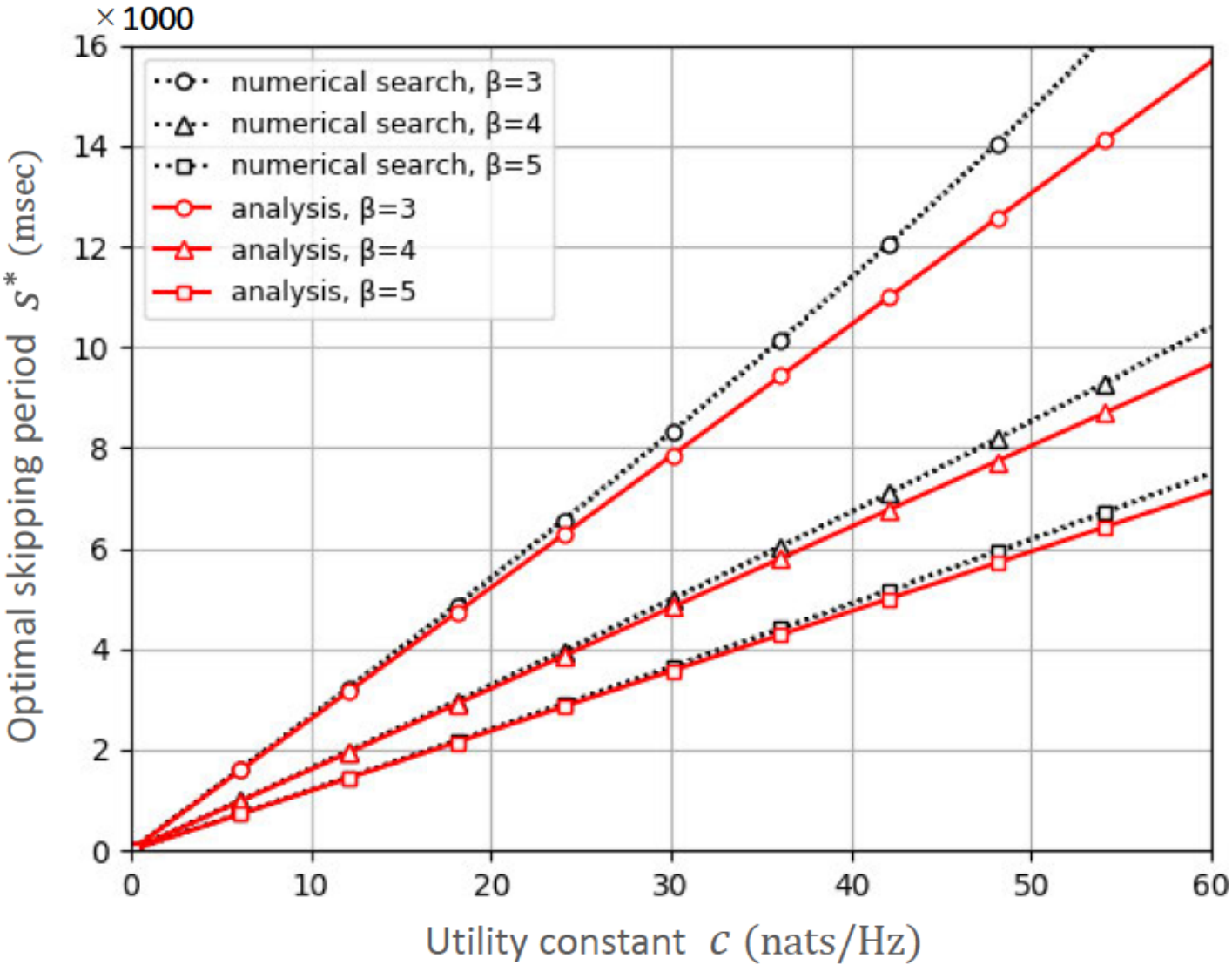}%
    \label{subfig:vsc}%
  }%
  \caption{The values of $s^*$ in \eqref{eq:s_opt} with respect to the
    path-loss exponent~$\beta$ and the utility constant~$c$ in
    \eqref{eq:U}.}%
  \label{fig:s_opt-vs-beta&c}
\end{figure}

We here numerically evaluate $s^*$ obtained in
Theorem~\ref{thm:s_opt}.
Figure~\ref{fig:s_opt-vs-beta&c} shows the numerical results of $s^*$
(in $1\text{slot}=1\text{msec}$) with respect to the path-loss
exponent~$\beta$ and the utility constant~$c$.
The BS intensity and the moving speed of the typical UE are fixed as
$\lambda=1$~(units/km$^2$) and $v=0.01$~(km/sec), respectively.
For comparison, the values of $s$, which locally maximize the lower
bound of the utility metric, obtained by replacing $\tau$ in
\eqref{eq:DataRate1} with its lower bound given in
\eqref{eq:tau1_lower}, are numerically searched and plotted in the
figure.
From Fig.~\ref{fig:s_opt-vs-beta&c}, we observe that the values of
$s^*$ agree well with the values from the numerical search even for
positive moving speed, in particular for large $\beta$ and small $c$,
in spite that $s^*$ is derived under the condition of sufficiently
small moving speed.
Figure~\ref{subfig:vsbeta} shows that $s^*$ is decreasing in $\beta$.
This is thought to be because $\tau(u)$ decays more rapidly with
respect to $u$ when $\beta$ is larger (as confirmed in
Fig.~\ref{fig:cmprsn_tau}); that is, smaller $s$ brings better
performance when $\beta$ is larger since $\mathcal{T}(s) =
\sum_{t=0}^{s-1}\tau(t v)/s$.
On the other hand, Fig.~\ref{subfig:vsc} shows that $s^*$ is linearly
increasing in $c$ (as is clear from \eqref{eq:s_opt}).
This is interpreted as because the utility metric in \eqref{eq:U} is
linearly decreasing in $c$, and thus better performance is given by
larger $s$ which makes the HO rate lower.

\begin{figure}[!t]
  \centering
  \subfloat[$s^*$ as a function of $v$ with several patterns of
    $\lambda$ and $\beta$, where $c=10$~(nats/Hz) is fixed.]{%
    \includegraphics[width=.5\linewidth]{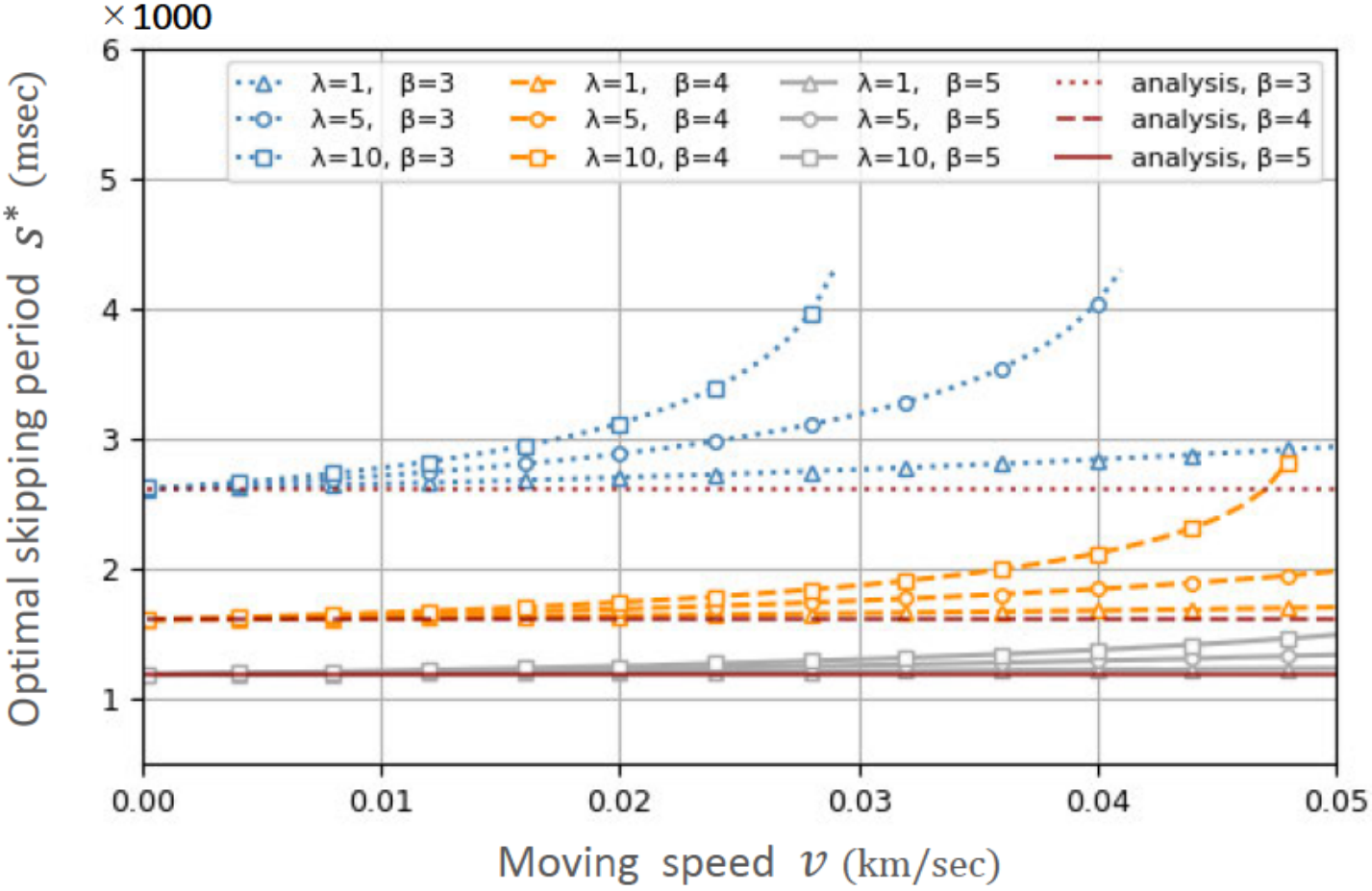}%
    \label{subfig:varyingbeta}%
  }
  \\
  \subfloat[$s^*$ as a function of $v$ with several patterns of
    $\lambda$ and $c$, where $\beta=5$ is fixed.]{%
    \includegraphics[width=.5\linewidth]{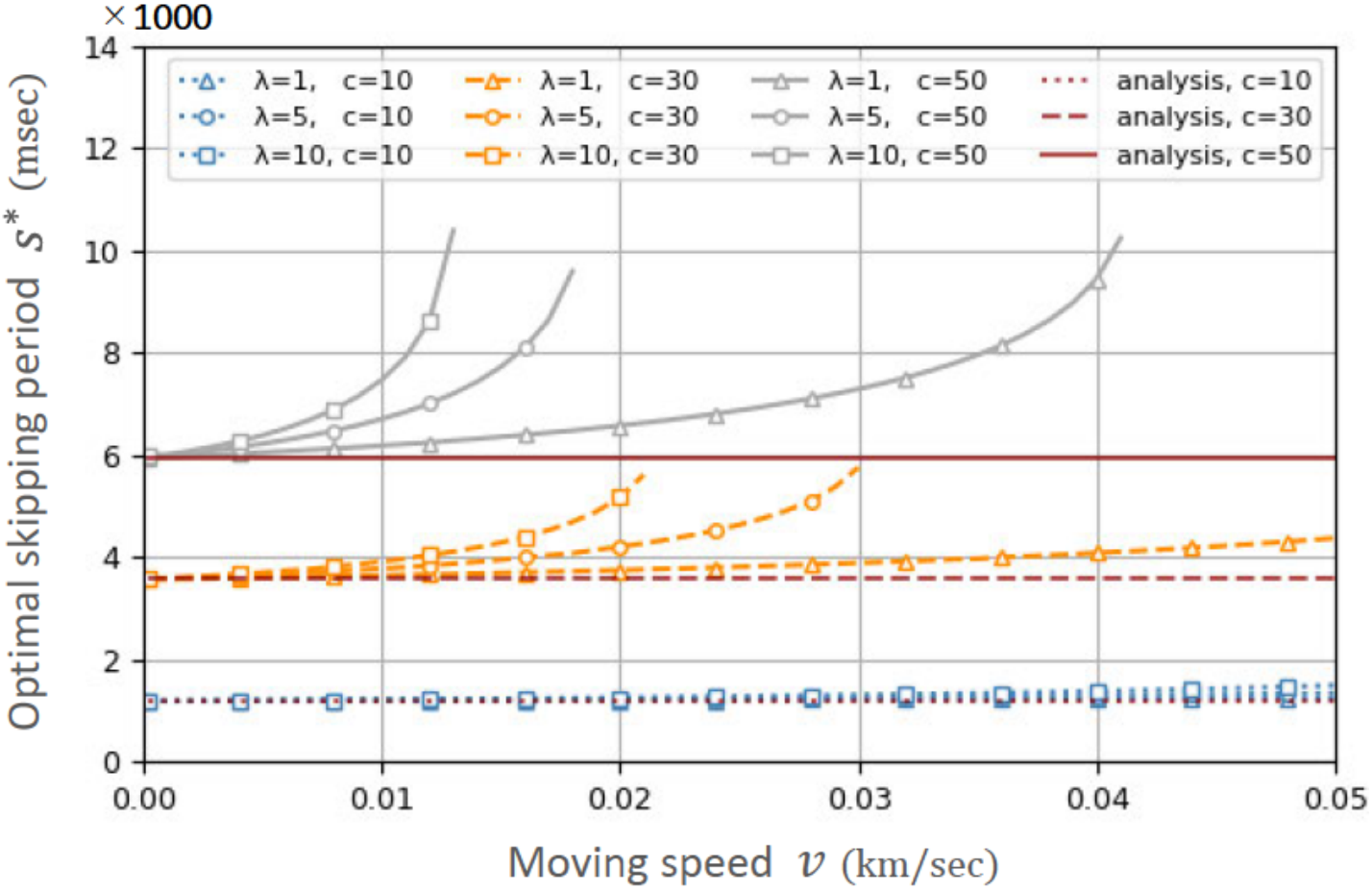}%
    \label{subfig:varyingc}%
  }%
  \caption{The values of $s^*$ in \eqref{eq:s_opt} with respect to the
    moving speed~$v$ of the typical UE.}%
  \label{fig:s_opt-vs-v}
\end{figure}

Figure~\ref{fig:s_opt-vs-v} further compares the values of $s^*$ in
\eqref{eq:s_opt} and the numerically searched values of $s$ as above
with respect to the constant moving speed~$v$ of the typical UE.
Since $s^*$ does not depend on $v$ and $\lambda$ (see
\eqref{eq:s_opt}), its value is given as a horizontal line for each
pair of $\beta$ and $c$.
Note that the values of the optimal skipping period obtained by the
numerical search do not change significantly with respect to the
changes in $v$ and $\lambda$, in particular for large $\beta$ and
small $c$, which allows us to use $s^*$ in \eqref{eq:s_opt} as an
approximation of the optimal skipping period for any $v$ and
$\lambda$, in particular when $\beta$ is large and $c$ is small.

\section{Comparison with Other Techniques}\label{sec:comparison}

\begin{figure}[!t]
  \centering
  \includegraphics[width=.5\linewidth]{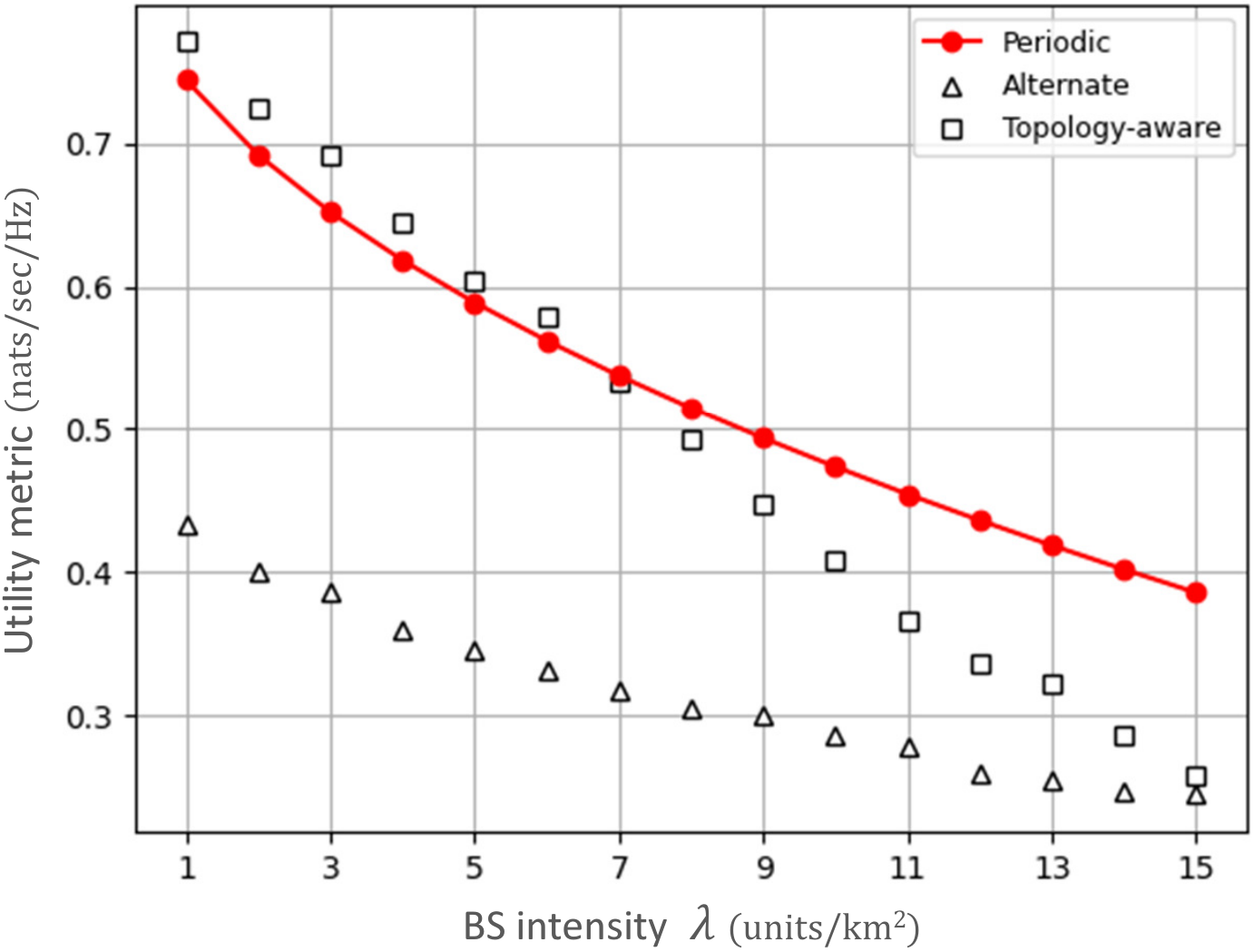}
  \caption{Numerical comparison of the utility metrics for the
    periodic HO skipping and other related HO skipping techniques.}%
  \label{fig:Compare_Schemes}
\end{figure}

We now compare our proposed scheme with two other related HO skipping
techniques---the alternate HO skipping
in~\cite{ArshElSaSoroAlNaAlou16} and the topology-aware HO skipping
in~\cite{DemaPsomKrik18}.
We use the utility metric~$\mathcal{U}$ in \eqref{eq:U} for the
comparison study.
In the computation of $\mathcal{U}$ in our proposed scheme, we adopt
the approximate optimal skipping period obtained in
Theorem~\ref{thm:s_opt}, and in the computation of the expected
downlink data rate, $\tau$ in \eqref{eq:DataRate1} is replaced by its
approximation~\eqref{eq:tau1_approx} with the adjustment function
$\epsilon(u)= e^{-10\,u^2}$, $u\ge0$.
In the other two comparison techniques, the expected downlink data
rates and the HO rates are computed using Monte Carlo simulation.
The numerical results are shown in Fig.~\ref{fig:Compare_Schemes},
where we set as $1\text{slot}=1\text{msec}$ and the values of the
utility metrics with respect to the BS intensity~$\lambda$ are
plotted.
In the topology-aware HO skipping, the chord length threshold is fixed
as $M=0.3$~(km) (see~\cite{DemaPsomKrik18} for details).
The other parameters are set as $\beta=3$, $c=10$~(nats/Hz), and the
moving speed is constant as $v=0.01$~(km/sec).
From the figure, we observe that our proposed scheme can outperform
the other two techniques depending on the parameter setting.
Although it is not possible to examine all the combinations of
parameter setting, we could at least assert that our proposed scheme
is comparable to the other sophisticated techniques in spite of its
simpleness.

\section{Conclusion}\label{sec:conclusion}

In this paper, we have proposed a simple HO skipping scheme in
cellular networks, called the periodic HO skipping, and have evaluated
its performance analytically and numerically.
Specifically, applying the stochastic geometry approach, we have
derived numerically computable expressions for the expected downlink
data rate and the HO rate when the UE adopts the proposed scheme.
Through the numerical experiments based on the analysis, we have
confirmed that the proposed scheme can outperform the scenario without
any HO skipping in terms of a utility metric representing the
trade-off between the expected downlink data rate and the HO rate, in
particular when the UE moves fast.
Moreover, we have discussed how to decide the length of the skipping
period and have provided a simple computable expression of the
skipping period which approximately gives a local maximum of the
utility metric.
Numerical comparison with other related HO skipping techniques have
also shown that the proposed scheme is comparable to the others.
Although we have considered here a simple mobility model on a
homogeneous PPP network model, further development within more
extended and generalized frameworks (e.g., HetNets with interference
cancellation and/or the BS cooperation) would be expected for future
work and one direction of the extensions is found in
\cite{XuTokuWada22}.

\appendices
\section{Proof of Lemma~\ref{lem:tau1}}\label{app:Prf_Lemma_tau1}

Applying Hamdi's Lemma~\cite[Lemma~1]{Hamd10} to the expectation of
\eqref{eq:DataRate} with \eqref{eq:SINR}, \eqref{eq:Interference} and
$i=B(\bm{0})$, we have
\begin{align}\label{eq:prf_lem_tau1}
  \Exp[\xi_{\bm{u},B(\bm{0})}(t)]
  &= \Exp\Biggl[
       \log\Biggl(
         1 + \frac{H_{B(\bm{0}),t}\,\|X_{B(\bm{0})} - \bm{u}\|^{-\beta}}
                  {I_{\bm{u},B(\bm{0})}(t) + \sigma^2}
       \Biggr)
     \Biggr]      
  \nonumber\\    
  &= \int_0^\infty
       \frac{e^{-\sigma^2 z}}{z}\,
       \Bigl(
         \Exp\bigl[
           e^{-z\, I_{\bm{u},B(\bm{0})}(t)}
         \bigr]
         - \Exp\bigl[
             e^{-z\sum_{j\in\N}H_{j,t}\,\|X_j - \bm{u}\|^{-\beta}}
           \bigr]
       \Bigr)\,
     \dd z.
\end{align}
For the second expectation in the last expression above, the Laplace
transform of an exponential distribution and the generating functional
of a PPP~(e.g., \cite[Example~9.4(c)]{DaleVere08}) yield
\begin{align}\label{eq:prf_lem_tau3}
  \Exp\Biggl[
    \prod_{j\in\N}e^{-z\,H_{j,t}\,\|X_j - \bm{u}\|^{-\beta}}
  \Biggr]
  &= \Exp\Biggl[
       \prod_{j\in\N}
         \Bigl(1 + \frac{z}{\|X_j - \bm{u}\|^\beta}\Bigr)^{-1}
     \Biggr]
  \nonumber\\
  &= \exp\biggl(
       - \lambda\,z
         \int_{\R^2}
           \frac{1}{z + \|\bm{x}\|^\beta}\,
         \dd\bm{x}
     \biggr)
  \nonumber\\
  &= e^{-\pi\lambda K_\beta\,z^{2/\beta}},
\end{align}
with $K_\beta = (2\pi/\beta)\csc(2\pi/\beta)$ as given in
\eqref{eq:K_beta}, where we use the polar coordinate conversion and
$\int_0^\infty w^{a-1}/(1+w)\,\dd w = \pi\csc a\pi$ for $a\in(0,1)$ in
the last equality.
Next, we consider the first expectation in the last expression of
\eqref{eq:prf_lem_tau1}, which satisfies
\begin{align}\label{eq:prf_lem_tau2}
  \Exp\bigl[
    e^{-z\,I_{\bm{u},B(\bm{0})}(t)}
  \bigr]
  &= \int_0^\infty
       \Exp\bigl[
         e^{-z\,I_{\bm{u},B(\bm{0})}(t)}
       \bigm|
         \|X_{B(\bm{0})}\| = r  
       \bigr]\,
       f_0(r)\,
     \dd r,
\end{align}
where $f_0(r) = 2\pi\lambda\,r e^{-\pi\lambda r^2}$ gives the
probability density function of $\|X_{B(\bm{0})}\|$.
Similar to obtaining \eqref{eq:prf_lem_tau3}, we have
\begin{align}\label{eq:prf_lem_tau6}
  &\Exp\bigl[
     e^{-z\,I_{\bm{u},B(\bm{0})}(t)}
   \bigm|
     \|X_{B(\bm{0})}\| = r  
   \bigr]
  \nonumber\\
  &= \Exp\Biggl[
       \prod_{j\in\N\setminus\{B(\bm{0})\}}
         \Bigl(1 + \frac{z}{\|X_j - \bm{u}\|^\beta}\Bigr)^{-1}
     \Biggm|
       \|X_{B(\bm{0})}\| = r  
     \Biggr]
  \nonumber\\
  &= \exp\biggl(
       - \lambda\,z
         \int_{\|\bm{x}\|>r}
           \frac{1}{z + \|\bm{x}-\bm{u}\|^\beta}\,
         \dd\bm{x}
     \biggr)
  \nonumber\\
  &= \exp\biggl(
       - \pi\lambda K_\beta\,z^{2/\beta}  
       + \lambda\,z
         \int_{\|\bm{x}\|\le r}
           \frac{1}{z + \|\bm{x}-\bm{u}\|^\beta}\,
         \dd\bm{x}
     \biggr),
\end{align}
where the polar coordinate conversion gives
\begin{equation}\label{eq:prf_lem_tau5}
  z\int_{\|\bm{x}\|\le r}
    \frac{1}{z + \|\bm{x}-\bm{u}\|^\beta}\,
  \dd\bm{x}
  = J(r,z,u),
\end{equation}
with $J$ given in \eqref{eq:J1}.
Plugging \eqref{eq:prf_lem_tau6} together with \eqref{eq:prf_lem_tau5}
into \eqref{eq:prf_lem_tau2}, we have
\begin{equation}\label{eq:prf_lem_tau4}
  \Exp\bigl[
    e^{-z\,I_{\bm{u},B(\bm{0})}(t)}
  \bigr]
  = e^{-\pi\lambda K_\beta\,z^{2/\beta}}\,
    \mu(z,u),
\end{equation}
with $\mu$ in~\eqref{eq:rho1}.
Finally, plugging \eqref{eq:prf_lem_tau3} and \eqref{eq:prf_lem_tau4}
into \eqref{eq:prf_lem_tau1} derives \eqref{eq:tau1}.

\section{Proof of Lemma~\ref{lem:J1}}\label{app:Prf_Lem_J1}

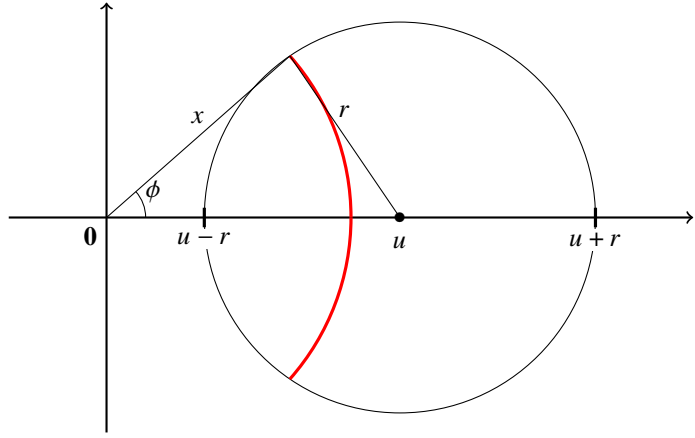
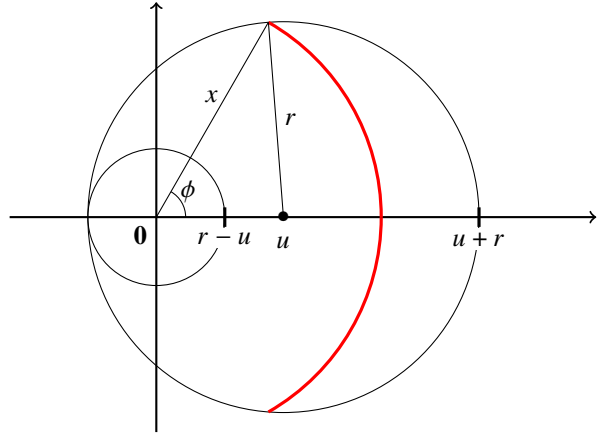
\begin{figure}[!t]
  \small\centering
  \subfloat[Case of $u\ge r$, where $x$ varies from $u-r$ to $u+r$.]{%
  \begin{tikzpicture}[scale=1.3]
    \draw[thick,->](-1,0)--(6,0);
    \draw[thick,->](0,-2.2)--(0,2.2);
    \draw(0,0) node[below left]{$\bm{0}$};
    \draw(3,0) circle (2);
    \draw(3,0) node{$\bullet$};
    \draw(3,-.1) node[below]{$u$};
    \draw[very thick, color=red](2.5,0) arc (0:41.40962212:2.5);
    \draw[very thick, color=red](2.5,0) arc (0:-41.40962212:2.5);
    \draw(0,0)-- node[above=1.5pt]{$x$}(1.875,1.65359456942);
    \draw(3,0)-- node[above=4pt]{$r$}(1.875,1.65359456942);
    \draw[very thick](1,.1)--(1,-.1)
      node[below,fill=white,inner sep=1.5pt]{$u-r$};
    \draw[very thick](5,.1)--(5,-.1)
      node[below,fill=white,inner sep=1.5pt]{$u+r$};
    \draw[thin](.4,0) arc (0:41.40962212:.4) node[right]{$\phi$};
  \end{tikzpicture}%
  \label{subfig:u>r}%
  }%
  \\
  \subfloat[Case of $u < r$, where $x$ varies from $r-u$ to $u+r$.]{%
  \begin{tikzpicture}[scale=1.3]
    \draw[thick,->](-1.5,0)--(4.5,0);
    \draw[thick,->](0,-2.2)--(0,2.2);
    \draw(0,0) node[below left]{$\bm{0}$};
    \draw(1.3,0) circle (2);
    \draw(1.3,0) node{$\bullet$};
    \draw(1.3,-.1) node[below]{$u$};
    \draw(0,0) circle (.7);
    \draw[very thick, color=red](2.3,0) arc (0:60.1105727645:2.3);
    \draw[very thick, color=red](2.3,0) arc (0:-60.1105727645:2.3);
    \draw(0,0)-- node[above=3pt]{$x$}(1.14615384615,1.99407406105);
    \draw(1.3,0)-- node[right]{$r$}(1.14615384615,1.99407406105);
    \draw[very thick](.7,.1)--(.7,-.1)
      node[below,fill=white,inner sep=1.5pt]{$r-u$};
    \draw[very thick](3.3,.1)--(3.3,-.1)
      node[below,fill=white,inner sep=1.5pt]{$u+r$};
    \draw[thin](.3,0) arc (0:60.1105727645:.3) node[right]{$\phi$};
  \end{tikzpicture}%
  \label{subfig:u<r}%
  }%
  \caption{Supplement to the derivation of eqs.~\eqref{eq:prf_lem_J1}
    and \eqref{eq:prf_lem_J2}, where
    $\phi=\disp{\arccos\Bigl(\frac{x^2+u^2-r^2}{2x u}\Bigr)}$ for each
    $x\in[|u-r|,u+r]$.}%
  \label{fig:suppl}
\end{figure}

It is immediate for the case of $u=0$ since $w_{x,0,\phi}=x$ in
\eqref{eq:J1}.
Suppose $u>0$.
On the left-hand side of \eqref{eq:prf_lem_tau5}, changing the
variables as $\bm{x}' = \bm{u}-\bm{x}$ leads to
\[
  \int_{\|\bm{x}\|\le r}
    \frac{1}{z + \|\bm{x}-\bm{u}\|^\beta}\,
  \dd\bm{x}
  = \int_{b_{\bm{u}}(r)}
      \frac{1}{z + \|\bm{x}'\|^\beta}\,
    \dd\bm{x}',
\]    
where $b_{\bm{u}}(r)$ denotes the disk centered at $\bm{u}\in\R^2$
with radius~$r>0$.
Recall that $\|\bm{u}\|=u$ as in Lemma~\ref{lem:tau1}.
When $u\ge r$, the polar coordinate conversion gives (see
Fig.~\ref{subfig:u>r})
\begin{align}\label{eq:prf_lem_J1}
  &\int_{b_{\bm{u}}(r)}
     \frac{1}{z + \|\bm{x}\|^\beta}\,
   \dd\bm{x}
  \nonumber\\
  &= 2\int_{u-r}^{u+r}
        \frac{x}{z+x^\beta}\,
        \arccos\Bigl(\frac{x^2+u^2-r^2}{2x u}\Bigr)\,
      \dd x
  \nonumber\\
  &= 2\int_0^{u+r}
        \frac{x}{z+x^\beta}\,
        \arccos\Bigl(\frac{x^2+u^2-r^2}{2x u}\wedge1\Bigr)\,
      \dd x,
\end{align}
where the last equality holds since $f(x)=(x^2+u^2-r^2)/(2x u) > 1$
for $x\in(0,u-r)$ with $f(u-r)=1$ when $u\ge r>0$.
On the other hand, when $u<r$, we have similarly (see
Fig.~\ref{subfig:u<r}),
\begin{align}\label{eq:prf_lem_J2}
  &\int_{b_{\bm{u}}(r)}
     \frac{1}{z + \|\bm{x}\|^\beta}\,
   \dd\bm{x}
  \nonumber\\
  &= 2\pi\int_0^{r-u}
        \frac{x}{z+x^\beta}\,
     \dd x
%  \nonumber\\
%  &\quad\mbox{}   
     + 2\int_{r-u}^{u+r}
        \frac{x}{z+x^\beta}\,
        \arccos\Bigl(\frac{x^2+u^2-r^2}{2x u}\Bigr)\,
      \dd x        
  \nonumber\\
  &= 2\int_0^{u+r}
        \frac{x}{z+x^\beta}\,
        \arccos\Bigl(-1\vee\frac{x^2+u^2-r^2}{2x u}\Bigr)\,
      \dd x,
\end{align}
where the last equality holds since $f(x)=(x^2+u^2-r^2)/(2x u)< -1$
for $x\in(0,r-u)$ with $f(r-u)=-1$ when $0<u<r$.
Hence, unifying \eqref{eq:prf_lem_J1} and \eqref{eq:prf_lem_J2},
we have \eqref{eq:J11} since $(x^2+u^2-r^2)/(2x u) \in[-1,1]$ when
$|u-r|\le x\le u+r$.

\section{Proof of Corollary~\ref{cor:tau1_lower}}\label{app:Prf_Cor_tau1}

In \eqref{eq:prf_lem_tau1} with $\sigma^2=0$, changing the variables
as $z'=\|X_{B(\bm{0})}-\bm{u}\|^{-\beta}\,z$ leads to
\begin{align}\label{eq:prf_cor_tau1_lower1}
  &\Exp[\xi_{\bm{u},B(\bm{0})}(t)]
  \nonumber\\
  &= \int_0^\infty
       \frac{1}{z'}\,
       \Exp\Bigl[
         e^{- \|X_{B(\bm{0})}-\bm{u}\|^\beta\,z'\,
             I_{\bm{u},B(\bm{0})}(t)}\,
         \bigl(1-e^{-z' H_{B(\bm{0}),t}}\bigr)
       \Bigr]\,
     \dd z'
  \nonumber\\
  &= \int_0^\infty
       \frac{1}{1+z}\,
       \Exp\Bigl[
         e^{- \|X_{B(\bm{0})}-\bm{u}\|^\beta\,z\, 
             I_{\bm{u},B(\bm{0})}(t)}
       \Bigr]\,
     \dd z,
\end{align}
where the second equality follows from
$\Exp[e^{-z\,H_{B(\bm{0}),t}}\mid B(\bm{0})] = (1+z)^{-1}$ because
$H_{i,t}$, $i\in\N$, $t\in\N_0$, are mutually independent and
exponentially distributed with unit mean.
Furthermore, since $\|X_{B(\bm{0})}\|$ follows the probability density
function~$f_0(r) = 2\pi\lambda\,r e^{-\pi\lambda r^2}$ and the angle
between $X_{B(\bm{0})}$ and $\bm{u}$ is uniformly distributed on
$[0,2\pi)$, the expectation in the last expression of
\eqref{eq:prf_cor_tau1_lower1} satisfies
\begin{align}\label{eq:prf_cor_tau1_lower2}
  &\Exp\Bigl[
    e^{- \|X_{B(\bm{0})}-\bm{u}\|^\beta\,z\,
        I_{\bm{u},B(\bm{0})}(t)}
   \Bigr]
  \nonumber\\
  &= \frac{1}{2\pi}
     \int_0^{2\pi}\!\!\!\int_0^\infty
       \Exp\Bigl[
         e^{- {w_{r,u,\phi}}^\beta\,z\,I_{\bm{u},B(\bm{0})}(t)}
       \Bigm|
         \|X_{B(\bm{0})}\|=r
       \Bigr]\,
       f_0(r)\,
     \dd r\,\dd\phi
  \nonumber\\
  &\ge \lambda
       \int_0^{2\pi}\!\!\!\int_0^\infty
         r\,\exp\Bigl(
           -\pi\lambda r^2
           - \pi\lambda\,K_\beta\,{w_{r,u,\phi}}^2\,z^{2/\beta}
         \Bigr)\,
       \dd r\,\dd\phi
  \nonumber\\
  &= \lambda
      e^{-\pi\lambda K_\beta\,u^2\,z^{2/\beta}}
      \int_0^{2\pi}\!\!\!\int_0^\infty
        r\,\exp\Bigl(
          - \pi\lambda\bigl[
              (1+K_\beta\,z^{2/\beta})\,r^2
              - 2K_\beta\,u\,z^{2/\beta}r\cos\phi
            \bigr]
        \Bigr)\,
      \dd r\,\dd\phi
  \nonumber\\
  &= \frac{1}{1+K_\beta\,z^{2/\beta}}\,
     \exp\biggl(
       -\pi\lambda\,u^2\,
       \frac{\,K_\beta\,z^{2/\beta}}
            {1+K_\beta\,z^{2/\beta}}
     \biggr),
\end{align}  
where $w_{r,u,\phi} = \sqrt{r^2+u^2-2r u\cos\phi}$ and the inequality
follows from \eqref{eq:prf_lem_tau6}, from which the nonnegative
integral term is removed.
In the last equality in \eqref{eq:prf_cor_tau1_lower2}, we apply the
following; that is, for $p>0$ and $q\in\R$,
\begin{align*}
  \int_0^{2\pi}\!\!\!\int_0^\infty
    r\,e^{-p\,r^2 + q\,r\cos\phi}\,\dd r\,\dd\phi
  &= \int_{-\infty}^\infty\!\int_{-\infty}^\infty
       e^{-p(x^2+y^2) + q x}\,\dd x\,\dd y   
  \\
  &= e^{q^2/(4p)}
     \int_{-\infty}^\infty\!\int_{-\infty}^\infty
       e^{-p(x^2+y^2)}\,\dd x\,\dd y
  \\
  &= \frac{\pi}{p}\,e^{q^2/(4p)}.
\end{align*}
Finally, plugging \eqref{eq:prf_cor_tau1_lower2} into
\eqref{eq:prf_cor_tau1_lower1} and changing the variables as $z' =
K_\beta\,z^{2/\beta}$, we obtain \eqref{eq:tau1_lower}.

\section{Proof of Theorem~\ref{thm:s_opt}}
\label{app:Prf_Thm_s_opt}

Following the discussion prior to Theorem~\ref{thm:s_opt}, we
approximately derive the length of the skipping period that maximizes
the lower bound of the utility metric for sufficiently small moving
speed of the typical UE.
As in \eqref{eq:tau1_approx}, let $\Tilde{\tau}$ denote the lower
bound of $\tau$ given on the right-hand side of \eqref{eq:tau1_lower}.
As the first step of the approximation, we introduce continuous
relaxation that considers $s$ as a nonnegative real number though
the skipping period essentially takes an integer in our discrete-time
setting.
Furthermore, we replace the sum in \eqref{eq:DataRate1} with an
integral.
Namely, when the typical UE moves at constant speed~$v$, the lower
bound of the expected downlink data rate in \eqref{eq:DataRate1} is
approximated as
\begin{equation}\label{eq:dT/ds1}
  \Tilde{\mathcal{T}}(s)
  = \frac{1}{s}\int_{0}^s \Tilde{\tau}(t v)\,\dd t,
\end{equation}
which is now specified as a function of $s$.
Similarly, when the moving speed of the typical UE is constant at $v$,
the HO rate in \eqref{eq:HORate1} is reduced to
\begin{equation}\label{eq:dH/ds1}
  \mathcal{H}(s)
  = \frac{1}{s}\biggl[
      1 - 2\lambda
          \int_0^\pi\!\!\!\int_0^\infty
            r\,e^{- \lambda\,\eta(r, s v, \phi)}\,
          \dd r\,\dd\phi
    \biggr].
\end{equation}
We now consider the approximation of the lower bound of the utility
metric defined as
\begin{equation}\label{eq:U_tilde}
  \Tilde{\mathcal{U}}(s)
  = \Tilde{\mathcal{T}}(s) - c\,\mathcal{H}(s).
\end{equation}
Our purpose is then to approximately derive the solution~$s$ of the
equation that the following derivative is equal to $0$;
\begin{equation}\label{eq:dU/ds1}
  \frac{\dd\Tilde{\mathcal{U}}(s)}{\dd s}
  = \frac{\dd\Tilde{\mathcal{T}}(s)}{\dd s}
    - c\,\frac{\dd\mathcal{H}(s)}{\dd s}. 
\end{equation}
Consider the first term~$\dd\Tilde{\mathcal{T}}(s)/\dd s$ on the
right-hand side of \eqref{eq:dU/ds1}.
From~\eqref{eq:dT/ds1}, we have
\begin{equation}\label{eq:dT/ds2}
  \frac{\dd\Tilde{\mathcal{T}}(s)}{\dd s}
  = \frac{1}{s}\,\Tilde{\tau}(s v)
    - \frac{1}{s^2}\int_0^s \Tilde{\tau}(t v)\,\dd t.
\end{equation}
Taylor's theorem applied to the exponential in $\Tilde{\tau}$ on the
right-hand side of \eqref{eq:tau1_lower} leads to
\[
  \Tilde{\tau}(u)
  = \frac{\beta}{2}
    \int_0^\infty
      \frac{z^{\beta/2-1}}{(1+z)({K_\beta}^{\beta/2}+z^{\beta/2})}\,
      \Bigl(
        1 - \pi\lambda u^2 \frac{z}{1+z}
      \Bigr)\,
    \dd z
    + o(u^2)
    \quad\text{as $u\to0$.}
\]
Then, plugging this into \eqref{eq:dT/ds2}, we have for sufficiently
small $v>0$,
\begin{equation}\label{eq:dT/ds3}
  \frac{\dd\Tilde{\mathcal{T}}(s)}{\dd s}
  \approx
    - \frac{\pi\lambda\beta}{3}\,s v^2
      \int_0^\infty
        \frac{z^{\beta/2}}{(1+z)^2({K_\beta}^{\beta/2} + z^{\beta/2})}\,
      \dd z.
\end{equation}

Next, to consider $\dd\mathcal{H}(s)/\dd s$ in \eqref{eq:dU/ds1}, we
take the derivative of the integrand on the right-hand side of
\eqref{eq:dH/ds1}; that is,
\[
  \frac{\partial}{\partial s}
    r\,e^{- \lambda\,\eta(r, s v, \phi)}
  = - \lambda r\,
      e^{- \lambda\,\eta(r, s v, \phi)}
      \frac{\partial}{\partial s}\eta(r, s v, \phi),
\]
where \eqref{eq:eta} leads to
\[
  \frac{\partial}{\partial s}\eta(r, s v, \phi)
  = - 2\,v\Bigl[
        (r\cos\phi - s v)
        \arccos\Bigl(\frac{r\cos\phi - s v}{w_{r,sv,\phi}}\Bigr)
        - r\sin\phi
      \Bigr],
\]
with $w_{r,sv,\phi} = \sqrt{r^2 + (sv)^2 -2\,rsv\cos\phi}$.
By \eqref{eq:eta} and \eqref{eq:prf_prp_HO1}, we know that $\eta(r,
sv,\phi) = \pi r^2 + |b_{\bm{y}}(w_{r, sv,\phi})\setminus
b_{\bm{0}}(r)| \ge \pi r^2$, so that,
\[
  \biggl|
    \frac{\partial}{\partial s}
      r\,e^{- \lambda\,\eta(r, s v, \phi)}
  \biggr|
  \le 2\lambda v e^{-\pi\lambda r^2}\,
      \bigl[(\pi+1)r^2 + \pi s v r\bigr],
\]
and for any fixed $s\in(0,\infty)$ and $v\in(0,\infty)$,
\[
  \int_0^\pi\!\!\!\int_0^\infty
    \biggl|
      \frac{\partial}{\partial s}
        r\,e^{- \lambda\,\eta(r, s v, \phi)}
    \biggr|\,
  \dd r\,\dd\phi
  \le \frac{(\pi + 1)\,v}{2\lambda^{1/2}}
      + \pi s v^2 < \infty.
\]
Therefore, we can change the order of the integral and derivative, and
we have
\begin{align}\label{eq:dH/ds2}
  \frac{\dd\mathcal{H}(s)}{\dd s}
  &= \frac{2\lambda^2}{s}
     \int_0^\pi\!\!\!\int_0^\infty
       r\,e^{- \lambda\,\eta(r, s v, \phi)}\,
       \frac{\partial}{\partial s}\eta(r, s v, \phi)\,
     \dd r\,\dd\phi
  \nonumber\\
  &\quad\mbox{}        
   - \frac{1}{s^2}\biggl\{
       1 - 2\lambda
           \int_0^\pi\!\!\!\int_0^\infty
             r\,e^{- \lambda\,\eta(r, s v, \phi)}\,
           \dd r\,\dd\phi
     \biggr\}
  \nonumber\\
  &= \frac{2\lambda}{s}
     \int_0^\pi\!\!\!\int_0^\infty
       r\,e^{- \lambda\,\eta(r, s v, \phi)}\,    
       \biggl(
         \lambda\,\frac{\partial}{\partial s}\eta(r, s v, \phi)
         + \frac{1}{s}
       \biggr)\,
     \dd r\,\dd\phi
     - \frac{1}{s^2}.
\end{align}
Taylor's theorem applied to the integrand above gives
\begin{align*}
  &r\,e^{- \lambda\,\eta(r, s v, \phi)}\,    
   \Bigl(
     \lambda\,\frac{\partial}{\partial s}\eta(r, s v, \phi)
     + \frac{1}{s}
   \Bigr)
  \\
  &= r\,e^{-\pi\lambda r^2}\,
     \Bigl\{
       \frac{1}{s}
       - \lambda s v^2
         \bigl[
           2\lambda r^2 (\phi\cos\phi-\sin\phi)^2
           - \phi + \cos\phi\sin\phi
         \bigr]
     \Bigr\}    
     + o(v^2)
     \quad\text{as $v\to0$,}
\end{align*}
and we have for sufficiently small $v>0$,
\begin{align}\label{eq:dH/ds3}
  \frac{\dd\mathcal{H}(s)}{\dd s}
  &\approx
   \frac{2\lambda}{s}
   \int_0^\pi\!\!\!\int_0^\infty
     r\,e^{-\pi\lambda r^2}\,
       \Bigl\{
         \frac{1}{s}
         - \lambda s v^2
           \bigl[
             2\lambda r^2 (\phi\cos\phi-\sin\phi)^2
             - \phi + \cos\phi\sin\phi
           \bigr]
       \Bigr\}    
  \dd r\,\dd\phi     
  - \frac{1}{s^2}
  \nonumber\\
  &= - \Bigl(\frac{5}{2\pi} - \frac{\pi}{6}\Bigr)
       \lambda\,v^2.
\end{align}
Hence, plugging \eqref{eq:dT/ds3} and \eqref{eq:dH/ds3} into
\eqref{eq:dU/ds1} derives
\[
  \frac{\dd\Tilde{\mathcal{U}}(s)}{\dd s}
  \approx
  - \frac{\pi\lambda\beta}{3}\,s v^2
    \int_0^\infty
      \frac{z^{\beta/2}}{(1+z)^2({K_\beta}^{\beta/2} + z^{\beta/2})}\,
    \dd z  
  + \Bigl(\frac{5}{2\pi} - \frac{\pi}{6}\Bigr)
    c\,\lambda\,v^2.
\]
The right-hand side above is linearly decreasing in $s$, and solving
the equation that it is equal to $0$ with respect to $s$, we obtain
\eqref{eq:s_opt}.

\bibliographystyle{IEEEtran}
%\bibliography{tokuyama,../../references}%
\input{PeriodicHOSkipping.bbl}

\end{document}